\documentclass[12pt,preprint]{aastex}
\def \lp{\>\> .}
\def \lc{\>\> ,}

\def \arcmin{\hbox{$^{\prime}$}}

\def \cc{cm$^{-3}$}
\def \kms{km~s$^{-1}$}

\def \nh3{NH$_3$}
\def \n2h{N$_2$H$^+$}

\def \nh2{n_{H_2}}
\def \nh1{n_{HI}}

\def \tw{$^{12}$CO}
\def \th{$^{13}$CO}
\def \ce{C$^{18}$O}
\def \h2{H$_2$}

\def \c2{cm$^{-2}$}

\def \mic{$\mu$m}

\def \be{\begin{equation}}
\def \ee{\end{equation}}
\def \bf{\begin{figure}}
\def \ef{\end{figure}}

\def \lc{\>\> ,}
\def \lp{\>\> .}

\newcommand{\T}{\mathcal{T}}
\newcommand{\U}{\mathcal{U}}
\received{2004 May 13}
\begin{document}

\shorttitle{Further Studies of HI Narrow Self--Absorption}
\shortauthors{Goldsmith and Li}

\title{HI Narrow Self--Absorption in Dark Clouds:
  Correlations with Molecular Gas and Implications for Cloud Evolution and Star Formation}

\author{P. F. Goldsmith\altaffilmark{1} and D. Li\altaffilmark{2}}
\altaffiltext{1}{Department of 
Astronomy and National Astronomy and Ionosphere Center, Cornell University,\\ 
Ithaca NY 14853}

\altaffiltext{2}{Center for Astrophysics, 60 Garden Street, 
Cambridge MA 02138, dli@cfa.harvard.edu }

\begin{abstract}

We present the results of a comparative study of HI narrow self--absorption (HINSA), OH, 
\th, and \ce\ in five dark clouds.  
We find that the HINSA generally follows the distribution of the emission of the carbon
monoxide isotopologues, and has a characteristic size close to that of \th.  
This confirms earlier work \citep{li2003} which determined that the HINSA is produced by cold
HI which is well mixed with molecular gas in well--shielded regions.
The OH and \th\ column densities are essentially uncorrelated for the sources
other than L1544.
Our observations indicate that the central number densities of HI are between 2 and 6 cm$^{-3}$,
and that the ratio of the hydrogen density to total proton density for these sources is
5 to 27$\times10^{-4}$. 
Using cloud temperatures and the density of atomic hydrogen, we set an upper limit to
the cosmic ray ionization rate of 10$^{-16}$ s$^{-1}$. 
We present an idealized model for HI to \h2\ conversion in well--shielded regions, 
which includes cosmic ray destruction of \h2\ and formation of this species on grain surfaces.  
We include the effect of a distribution of grain sizes, and find that for a MRN 
distribution, the rate of \h2\ formation is increased by a factor of 3.4.

Comparison of observed and modeled fractional HI abundances indicates ages for these
clouds, defined as the time since initiation of HI $\rightarrow$ \h2\ conversion, to be
10$^{6.5}$ to 10$^{7}$ yr.  
Several effects may make this time a lower limit, but the low values of n$_{HI}$ we have
determined make it certain that the time scale for evolution from a possibly less dense
atomic phase to almost entirely molecular phase, must be a minimum of several million years.
This clearly sets a lower limit to the overall time scale for the process of star 
formation and the lifetime of molecular clouds.

\end{abstract}

\setcounter{footnote}{0}

\section{INTRODUCTION}

The study of atomic hydrogen in dense interstellar clouds has a long history.
Very early measurements of dark clouds indicated no correlation of atomic
hydrogen and dust opacity \citep{bok1955}.  Subsequent measurements in fact showed an
anticorrelation between the HI column density and extinction \citep[e.g.][]{garzoli1966}.
These pioneering observations of the Taurus region were followed by a broader
survey of dust clouds by \citet{heiles1969} in which a number of spectra showing 
``fluctuations'' were found, but one object showed clear evidence for ``self--absorption'': 
a narrow absorption feature superimposed on a wider emission line.  
In this source in Taurus, the velocity of the HI self--absorption agreed with
that of the OH emission.

Since that time, a number of large--scale and focused studies of HI in dark clouds
have been carried out \citep[e.g.][]{knapp1974, heiles1975, wilson1977, mccutcheon1978, 
myers1978, bowers1980, minn1981, batrla1981, shuter1987, vanderwerf1988, 
feldt1993, montgomery1995}.  
With improved sensitivity and higher angular and spectral resolution, most
of these studies identify self--absorption features.  
The inherent ambiguity between a temperature fluctuation in the emitting region and
a distinct, colder, foreground source responsible for the absorption makes it difficult to
distinguish between these two quite different situations for an arbitrary line 
of sight \citep[e.g.][]{gibson2000, kavars2003}, and
it is still only for relatively nearby clouds that the location of the cold atomic
hydrogen can be identified with certainty.

A key element in pinpointing the location of cold HI is its association with molecular emission,
and the greatest difference between early and current observations of HI self--absorption 
is the recognition that optically--identified dust clouds are cold (T $\simeq$ 10 K), 
dense (n $\ge$ 10$^3$ \cc)  primarily molecular regions.  
The availability of reasonably well--understood tracers which are surrogates for the
largely--unobservable \h2 means that determining the HI to \h2\ abundance ratio 
in nearby dark clouds is now practical.
These regions are of interest in their own right, as they contain the dense cores in which
low mass star formation can occur, and in regions such as the Taurus molecular cloud, these
are the only regions in which star formation appears to be occurring.
Additionally, because the visual extinction of these dark clouds is sufficient to make
photodestruction of \h2 unimportant in their central regions, they are ideal laboratories for
comparative studies of the atomic and molecular forms of hydrogen under relatively
simple conditions.

A recent survey of dark clouds using the Arecibo telescope revealed that over 80\% of the clouds
in Taurus exhibit HI narrow self--absorption \citep{li2003}, which we gave the acronym
{\it HINSA}.  
The atomic hydrogen responsible was shown to be associated with molecular material traced by OH,
\ce\, and \th\, on the basis of (1) the agreement in velocities, (2) the close agreement 
of nonthermal line widths, and (3) the low temperature implied by the depth of the HI features
and the small values of the nonthermal contribution to the line width.  Mapping data for only
one cloud was available, which showed reasonable agreement between the column densities of cold HI,
and OH, and the integrated intensity of \ce.

We have continued and extended this work, and in the present paper report observations of 
four clouds:  L1544, B227, L1574, and CB45.  
In Section \ref{observations} we present the observations of HI absorption, OH, \ce\, \th\, and
CI emission.
In Section \ref{colspecdist}, we discuss the spectra and determination of the column densities
of the species we have observed.  We also present the maps of these tracers in each of the sources.
In Section \ref{corr} we present and discuss correlations of the column densities of different
tracers.
In Section \ref{atmoldensities} we derive the number densities of the species and determine the
HI to \h2\ ratio for each of the clouds.  
In Section \ref{modeling} we present models for the time dependence of the atomic and molecular
hydrogen density.
Comparison of models and our data allows us to draw conclusions about the lifetime of these
clouds in Section \ref{discussion}.

\section{OBSERVATIONS}
\label{observations}

The HI and OH data were obtained using the 305 m radio telescope of the 
Arecibo Observatory\footnote{The Arecibo Observatory is part of the National
Astronomy and Ionosphere Center, which is operated by Cornell University
under a cooperative agreement with the National Science Foundation.}.
The data were obtained in four observing sessions between November 2002
and May 2003.  
The telescope was calibrated as described by \citet{li2003}
and both the HI and OH data, which were obtained simultaneously using the
``L-wide'' receiver, have been corrected for the main beam efficiency of 0.60.
The elliptical beam has FWHM beam dimensions of 3\arcmin.1 by 3\arcmin.6.
We observed using total power ``on source'' observations only, and removed
the receiver noise and system bandpass by fitting a low--order polynomial
to the spectrometer output.  
For these observations, the autocorrelation spectrometer
was configured to have three subcorrelators with 1024 lags each covering
0.391 MHz centered on frequencies of the HI, OH(1665 MHz) and OH(1667 MHz)
transitions.  
A fourth subcorrelator with 1024 lags covering a 0.781 MHz 
bandwidth was centered on the HI line, and was used for removing the instrumental
baseline when wider emission features were encountered.  
The nominal channel spacing for the HI observations was 0.08 \kms, but data
were typically smoothed to a velocity resolution of 0.16 \kms.  

The receiver temperature was typically 30 K at 1420, 1665, and 1667 MHz, 
but the total system temperature was more than double this value when observing
21--cm emission lines.  
The integration time per position was 1 minute.  This was sufficient to 
give good signal to noise for the HI absorption features, and also satisfactory 
OH 1667 MHz emission spectra.

Observations of \tw, \th, and \ce\ were obtained using the 
Five College Radio Astronomy Observatory 13.7 m radio telescope
\footnote{The Five College Radio Astronomy Observatory is operated by
the University of Massachusetts with support from the National Science
Foundation.}, during April and May 2003 and January 2004. 
The 32 pixel {\it Sequoia} receiver was configured to observe \th\ and \ce\
simultaneously for most of the data taking.  The autocorrelation spectrometer 
for each isotopologue had a bandwidth of 25 MHz and 1024 lags, corresponding
to a channel spacing in velocity of 0.067 \kms at 110 GHz.  
We mapped each source using ``on the fly'' data taking, covering a region 
determined by the HI observations obtained previously.  
The angular size of the maps ranged from 12\arcmin\ to 27\arcmin. 
System temperatures were typically 200 K for \th\ and \ce\, and 450 K
for \tw.
The integration time per sample point was nominally 10 s. 
The oversampled data was convolved to a circular Gaussian having the same 
solid angle as the Arecibo beam, corresponding to a FWHM equal to 3.3 \arcmin.  
We smoothed the data in velocity to the same 0.16 \kms\ resolution used for 
the HI and OH.  
Additional data on \tw\ for the source L1544 were obtained to be able to 
make accurate corrections for the optical depth of the \th\ emission.
All FCRAO data were corrected for appropriate main beam efficiencies, which are
0.49 for \th\ and \ce\, and 0.45 for \tw.

We obtained the data on atomic carbon (CI) using the Submillimeter Wave Astronomy
Satellite \citep[SWAS; ][]{melnick2000}.  Observations having total integration
time of 1 hour per position were obtained using position switching to a reference
position selected to have very weak or undetectable \tw\ J$\rightarrow$0 emission.
The data have been corrected for the main beam efficiency of 0.90.
The SWAS beam size is relatively large (3\arcmin.5  x 5\arcmin.0 at the 492 GHz
frequency of the CI line) and combined with low intensities of this transition in
dark clouds, we were only partially successful in being able to delineate the
distribution of CI relative to that of the other species studied.  
These results are further discussed in Section \ref{colspecdist}.

\section{COLUMN DENSITIES, SPECTRA, AND DISTRIBUTIONS}
\label{colspecdist}

Determination of the distances to dark clouds such as those in the present study presents
considerable difficulties.  L1544 is associated with the Taurus dark cloud, for which
the distance given here is the average of various techniques described by \citet{elias1978}.
B227 is at a galactic latitude of -0.46$^o$, and is located in a complex region of
considerable extinction. 
\citet{arquilla1984} observed the same cloud as studied here, and with a B-V color excess 
technique, derived a distance of 600 pc but there is obviously a large uncertainty.  
Tomita, Saito \& Ohtani (1979) studied the dark nebula L1570 
which is often \citep[e.g.][]{clemens1988} used interchangeably with B227, 
but there is a difference of over 8\arcmin\ between
the Arquilla and Tomita et al. positions.  
The latter authors, using star counts derived 
a distance 300 pc, but again with significant uncertainty.  
Earlier, \citet{bok1974} gave an ``adopted'' distance of 400 pc for B227.  
We here adopt a distance of 400 pc.
L1574 and CB45 are lumped together by \citet{kawamura1998} and are assigned
the 300 pc distance of the not too distant L1570 determined by \citet{tomita1979}, which
we adopt here.
The source coordinates and distances we have adopted are given in Table \ref{sources}.

\subsection{Cold HI}

The column density of the cold, relatively quiescent atomic hydrogen was
found by assuming an excitation temperature of 10 K, and solving for
the optical depth of the absorbing gas as described in \citet{li2003}.  
The maximum values of $\tau_{cold HI}$ are modest, ranging from 
0.34 for L1574r to 0.69 for L1574b.
As discussed in the earlier paper, the effect of foreground HI is primarily 
to reduce the apparent value of the 21--cm absorption optical depth.  
We have corrected the optical depth as described in \citet{li2003},
with values of p determined from the distances given in Table \ref{sources}.
The correction was less than 15 \%.
The corrected optical depth and the line width of the gaussian fit to the absorption profile
together yield the column density of cold absorbing gas using equation 18 
of \citet{li2003}. 

We present the peak main beam temperature T$_{mb}$ at the positions of maximum emission 
of each cloud in Table \ref{tpeak}. 
The values given are the peak antenna temperatures divided by
the main beam efficiency.

\subsection{OH}
The OH emission is almost certainly optically thin.  The peak main beam temperatures are 
typically 0.5 K, except in L1544 where the line is a factor of three stronger.  
The modest A--coefficient \citep[7.8$\times 10^{-11}$ s$^{-1}$; ][]{destombes1977} 
and typical collisional rate coefficients 
\citep[$\ge$ 2$\times10^{-11}$ cm$^3$ s$^{-1}$;][]{dewangan1987,
offer1994} result in a critical density $\leq$ 4 \cc, so that the excitation
temperature even in these cold clouds will be close to the kinetic temperature.  
We use equation 15 of \citet{li2003} to calculate the OH column densities 
from the integrated intensities.

\subsection{Carbon Monoxide Isotopologues}

The \ce\ 1-0 line is relatively easy to excite and the excitation temperature
of this line is again expected to be close to the kinetic temperature. 
In our sample, the \ce\
antenna temperatures are sufficiently small compared to the $\simeq$ 10 K 
kinetic temperatures to justify the assumption of low optical depth. 
We employ equation 18 of \citet{li2003} to derive the total column density,
assuming LTE and an excitation temperature of 10 K for all sources except L1544 
(disucssed below). 
The \th\ emission is also optically thin in all sources except L1544, where the 
positions of stronger emission near the center of the cloud have
optical depth of order unity.  
The comparison of \th\ and \ce\ emission (discussed further below) 
is consistent with the generally optically thick \th\ except in the
central region of L1544.

In order to assess and correct the situation for this cloud, we obtained a map of the
\tw\ emission from L1544.  
We derived the excitation temperature from the main beam temperature, assuming
that the emission is optically thick.  
Values for this source ranged from 9.5 K to
13 K, a few K warmer than typical for dark clouds. 
These excitation temperatures were used to compute the \ce\ column densities. 
We assumed that the \th\ has the same excitation temperature, and from its
main beam temperature, we derived the actual optical depth and corresponding column
density.  
The maximum ratio between the actual and the ``thin'' optical depths is a 
factor of 1.75, so the effect of the saturation on the \th\ is relatively modest and
confirms the appropriateness of assuming low optical depth for the \ce\ and for
the \th\ in clouds other than L1544.

In Table \ref{hioh1318} we give the column densities for the cold HI, OH, \th, and
\ce\ at the peak HINSA positions of the clouds observed.  
We also give the column density of \h2\ obtained from that of the carbon monoxide 
isotopologues using the expressions of \citet{frerking1982} for Taurus.
The molecular hydrogen column densities for each cloud derived using 
\th\ and \ce\ agree quite well.

\subsection{Uncertainties in Determination of the Column Densities}

The dominant contributions to the uncertainties in the integrated intensities of the 
various spectral lines observed in this study are quite different.
For the cold HI, this issue is discussed at length in \citet{li2003};   
for the HINSA, there are two major contributors.  The first is the foreground correction, which requires 
knowledge of the local HI distribution and the distance to the cloud for evaluation. 
This correction contributes at most 15\% for the present sample of clouds. 
The second contributor is the fitting of the background spectrum, which adds about 10\% for 
narrow absorption lines. 
The derived column density is inversely proportional to the assumed gas kinetic temperature.
The arguments in \citet{li2003} indicate that in these clouds, the HI responsible for the 
absorption is in cold gas traced by the carbon monoxide. 
The kinetic temperature is assumed to be 10 K, but the uncertainty is about 30\%.
The total uncertainty for HINSA is thus on the order of 50\%, with the statistical 
errors making a negligible contribution.

For the much weaker OH lines, statistical errors are more important, especially in the outer
portions of the clouds.  
The column density is only very weakly dependent on the kinetic temperature, and the emission
is almost certainly optically thin.
As discussed in \citep{li2003} ignoring the background can produce a systematic underestimate of
the OH column density by a factor $\simeq$ 2, which has not been corrected here.
For the carbon monoxide isotopologues, the uncertainty in the kinetic temperature dominates 
the determination of the column density (again except for the cloud edges).  
Assuming LTE, the column density changes by approximately 15\% if the kinetic temperature
varies from a nominal value of 10 K to 7 K or 15 K.  
As discussed above, for the central positions of L1544 the \th is slightly optically thick.  
The correction, which has been included, depends itself on the kinetic temperature, and
we estimate that this introduces an additional uncertainty of 25\%, but only for this
source.
We estimate that there is a factor of approximately 3 uncertainty in the \h2\
column densities due to variations in the fractional abundance of the carbon
monoxide isotopologues and uncorrected excitation and radiative transfer effects.

\subsection{Spectra}

In Figure \ref{spectra} we show the spectra of
four species at the reference position for each of the four clouds we have
studied.  
As discussed in detail by \citet{li2003}, the \ce\ and \th\ lines agree 
well in velocity and nonthermal line width with the OH emission and HINSA features. 
We here see, particularly evident for CB45, the close agreement of the complex absorption
profile of the HINSA with the profiles of the three emission lines observed in
the direction of this cloud.  This strongly supports the contention that the cold HI
is sampling the same region of space as the emission from the species studied here. 
The HI spectrum in the direction of B227 is complex, with various
peaks and dips.  Only the one of these, which is the narrowest and most
clearly defined absorption feature, is associated in velocity with detectable
molecular emission.  
This illustrates the difficulty of determining through HI spectroscopy alone whether 
any particular HI feature is a result of a variation in the temperature of
the emitting region or, if a dip, is due to absorption by cold gas in the
foreground.  

In the direction of L1574 there are two narrow HI absorption features;
each is quite well localized and has a very distinct morphology.  
The velocity of the ``blue'' feature shown here at the (0,9) 
\footnote{All offsets in
this paper are given as (minutes of arc in RA, minutes of arc in Decl.)}
position, the location 
of its maximum emission intensity, is 0 \kms, while the velocity of the ``red'' 
feature, which peaks at the (0,0) position, is 3.5 \kms.
In what follows we will refer to the cloud defined by the 0 \kms\ feature as
L1574b and that defined by the 3.5 \kms\ feature as L1574r.  
While the emission from L1574b overlaps that of L1574r, there is no clear indication
of any physical connection, so we consider these to be two independent clouds at
essentially the same distance along the line of sight.

\clearpage
\bf
\plotone{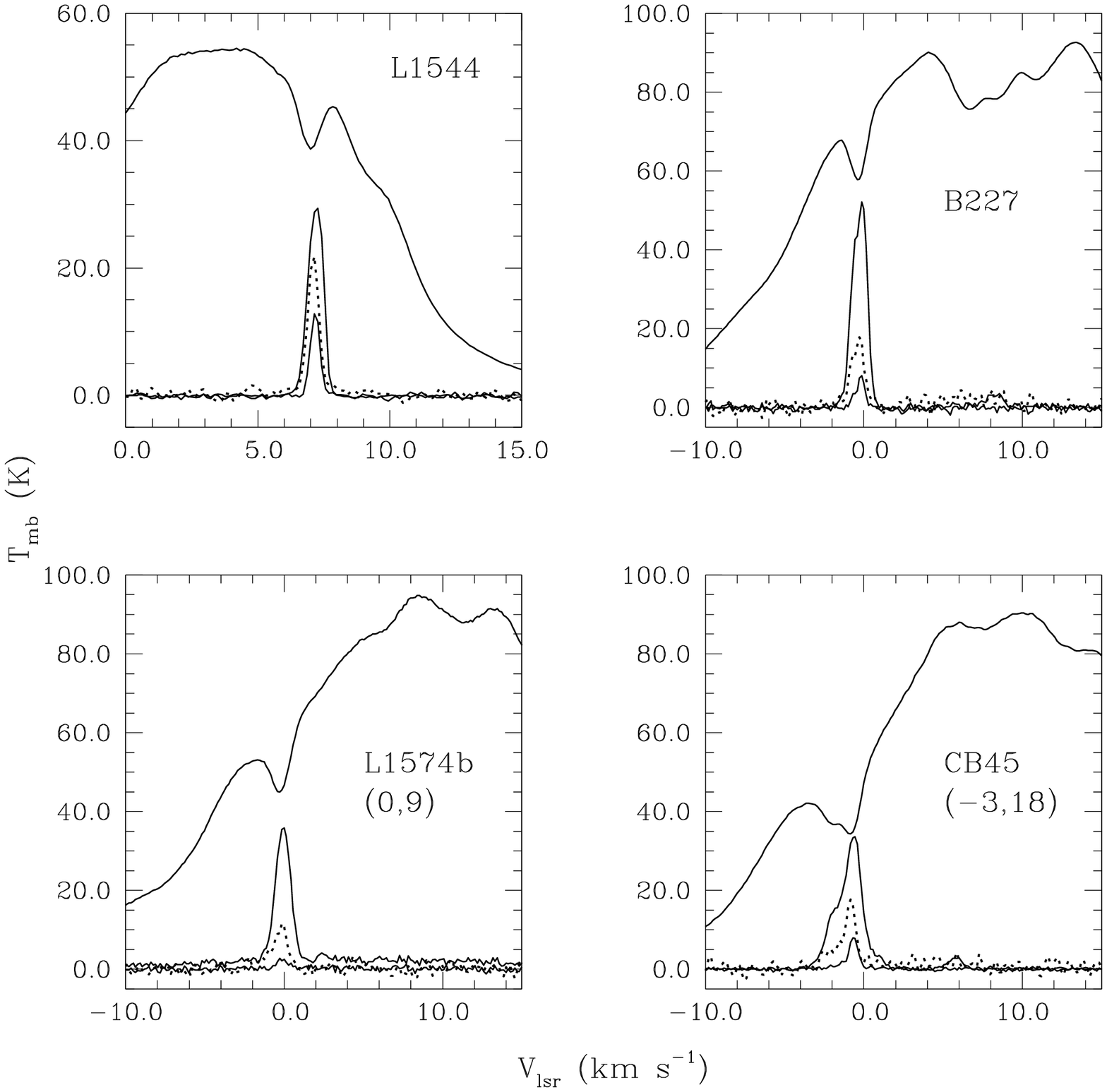}
\caption{\label{spectra} 
Spectra of different species in the four clouds included in this study.  
L1544 and B227 spectra are at the nominal central position of
each cloud while the offsets for L1574 and CB45 are in minutes of arc 
offset relative to the central position.
The offset position for L1574 corresponds to the peak of the 0 \kms\
feature and is denoted L1574b.
The spectra in for each position are arranged such that the HI lines,
having the greatest intensity, are plotted after scaling by factor 1.0.  
The \th\ lines are plotted after scaling by a factor of 5
for L1544, 20 for B227, 25 for L1574, and 10 for CB45, and are the
strongest after those of HI .
The OH(1667) spectra are plotted as dotted lines after scaling by factors 15 for L1544, 
50 for B227, 25 for L1574, and 40 for CB45, and fall below those of \th.  
The \ce\ lines are plotted after scaling 
by factors 5 for L1544, 20 for B227, 25 for L1574, and 10 for CB45,
and are the weakest spectral features.
}
\ef 
\clearpage

\subsection{Maps of Distributions}

In Figures \ref{l1544} through \ref{cb45} we show the HINSA cold--HI
column density (color) and \th\ and \ce\ column densities (contours) in the four
clouds.  
The millimeter data have been smoothed to approximate the spatial
resolution of the Arecibo telescope, as described previously.
Comparison of the distributions of the molecular emission, particularly the 
carbon monoxide isotopologues, reinforces the conclusion of \cite{li2003} that the
HI producing the narrow self--absorption features is in the cold, well--shielded
interiors of these molecular clouds.

In general the morphology of the molecular emission and HI absorption are quite 
similar.
This is particularly evident if one compares the atomic hydrogen with the \th\ in L1544 and L1574.  
In the latter cloud, we see very good agreement in both of the velocity components.  
The emission from CB45 peaks far from the (0,0) position, which is taken from \cite{clemens1988}.  
Systematic mapping around the nominal central position revealed the local maximum found within a
few arcminutes of the (0,0) position, but indicated an extension to the north.  
Following this up led to the far stronger maximum centered at the (-3,18) position,
which is close to the center of the HINSA absorption, although we see in Figure 
\ref{cb45} that the HI is concentrated in two features located on either side 
of the \ce\ emission maximum.

The exact peak of the \ce\ is offset from the strongest HINSA feature in two of the 
four sources, but only by a small fraction of the size of the cloud.
We thus feel that (1) the general distributions are very similar; (2) the strongest
molecular emission coincides quite closely with the strongest absorption by cold HI;
and (3) there is no indication of any limb brightening or other suggestion that the
cold HI producing the HINSA is not well mixed with the molecular gas. 
As discussed by \cite{li2003}, this may be due in part to the reduced absorption
coefficient of HI as the temperature increases, and thus we cannot exclude the
possibility that there is significant warm HI in an envelope or photon dominated
region (PDR) surrounding the molecular cloud.  However, based on comparison of 
line profiles and distributions of line intensities, the narrow line HI absorption
we are studying here is best interpreted as being produced by cold HI which is 
well mixed with the molecular gas.

\bf
\includegraphics[angle=270,scale=0.8]{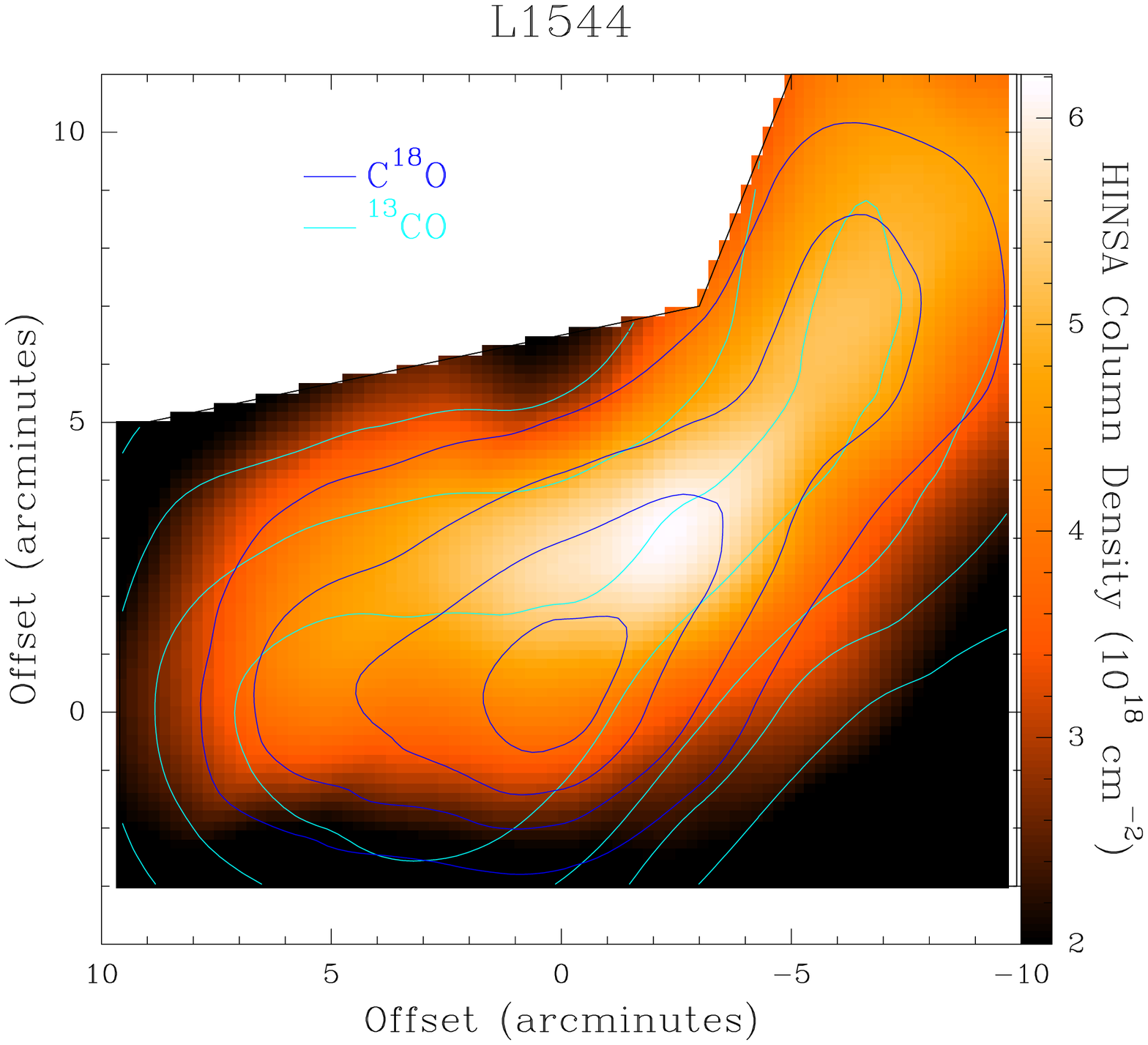}

\caption{\label{l1544} Maps of the cold HI column density producing 
narrow absorption features (HINSA in color; scale on right), contours of column densities
of \th\ (blue), and of \ce\ (magenta) in L1544.  
The contours are at 30, 50, 70, and 90\% of maximum values, which are 
$7.0\times10^{15}$  \c2\ for \th\ and $1.2\times10^{15}$ \c2\ for \ce.
}  
\ef

\bf
\includegraphics[scale=0.8]{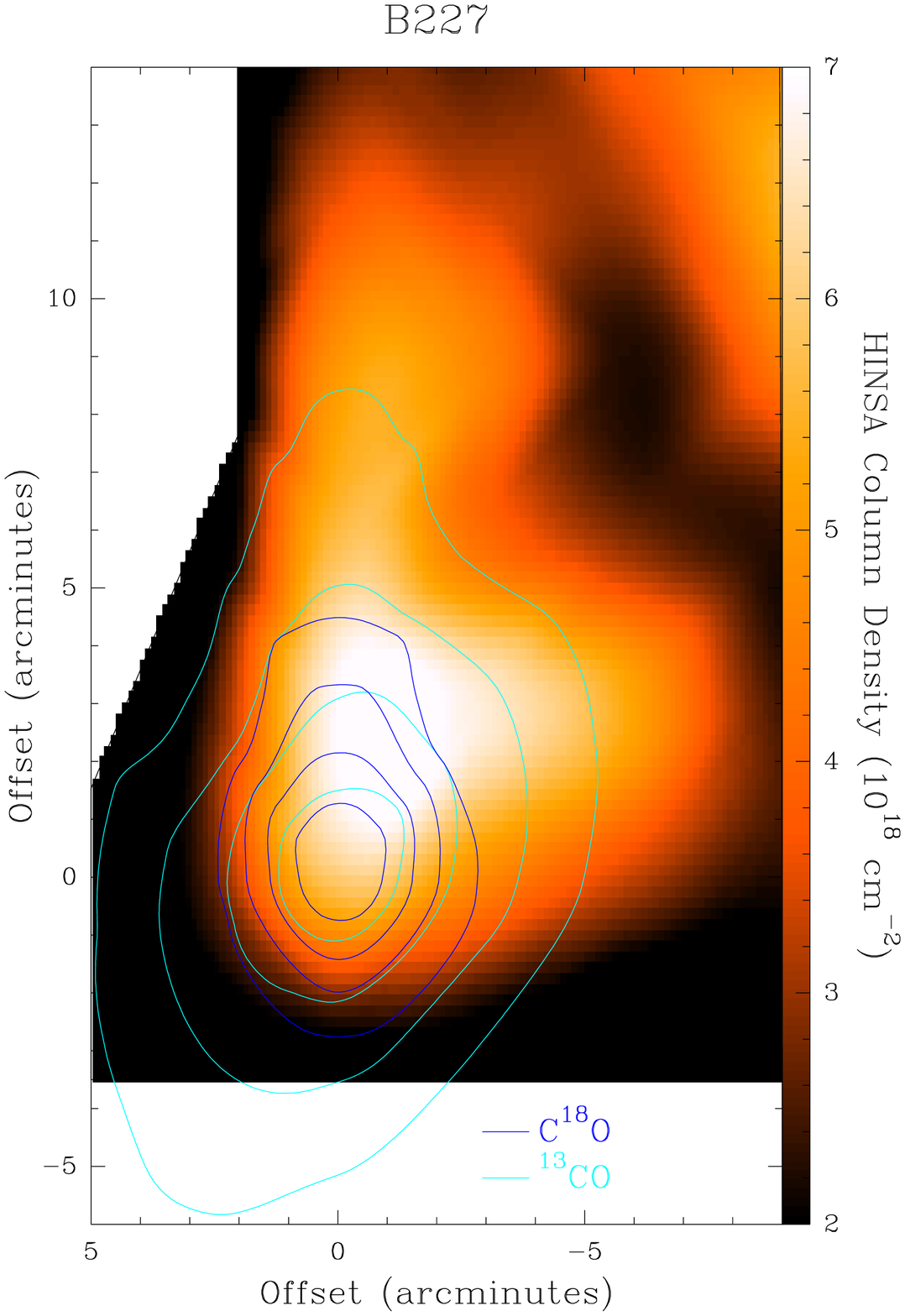}
\caption{\label{b227} HINSA, \th\ and \ce\ column densities for B227.
The color scale and relative contours are the same as in Figure \ref{l1544}.
The maximum value of the column density is $3.1\times10^{15}$  \c2\ for \th\ and $0.26\times10^{15}$ \c2\ for \ce.
}
\ef

\bf
\includegraphics[angle=270,scale=0.8]{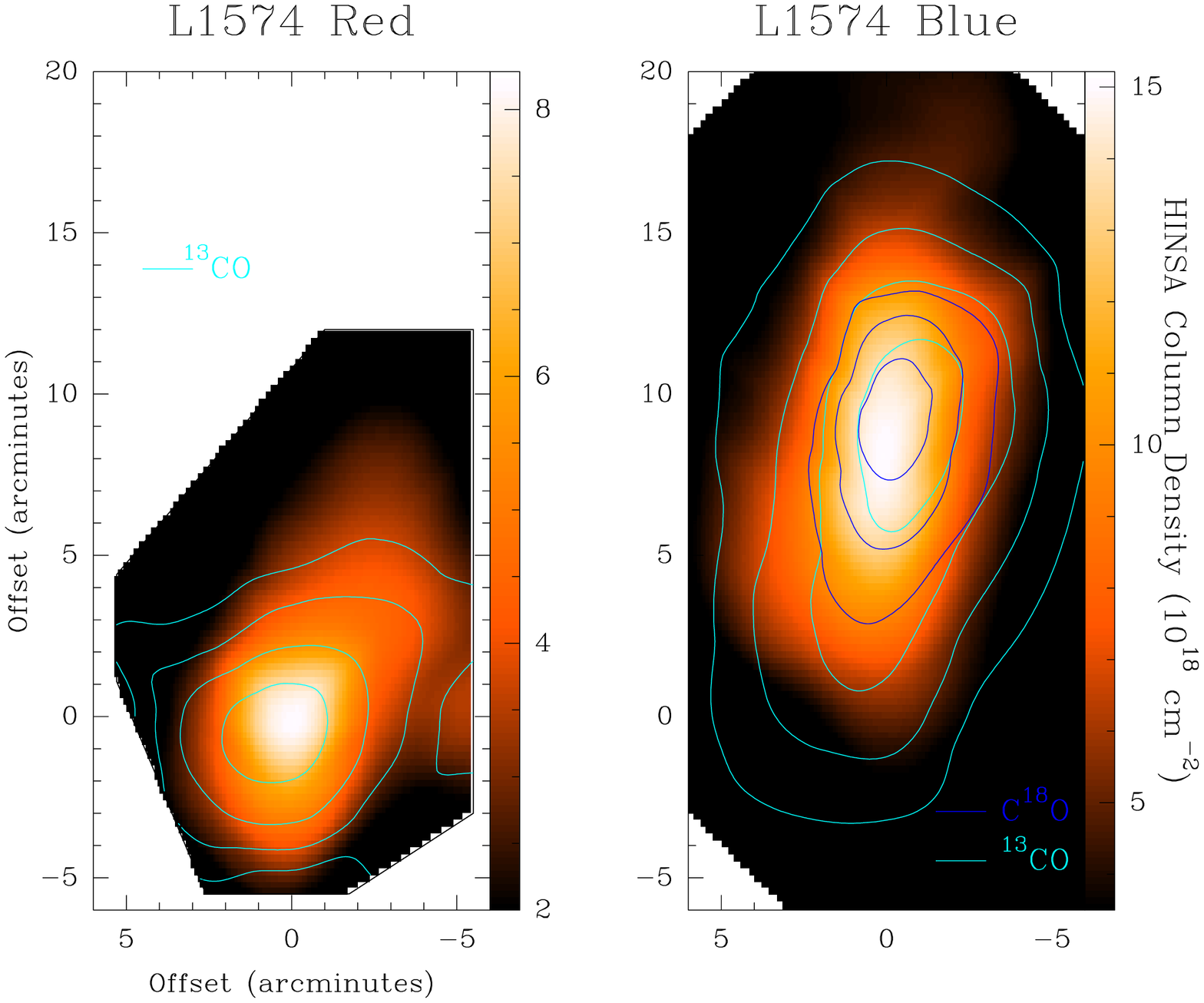}
\caption{\label{l1574}  HINSA, \th\ and \ce\ column densities for L1574.
The color scale and relative contours are the same as in Figure \ref{l1544}. 
For L1574b (velocity = 0 \kms), the maximum values of the column density are $3.6\times10^{15}$  
\c2\ for \th\ and $0.51\times10^{15}$ \c2\ for \ce. 
For L1574r (velocity = 3.5 \kms), the maximum \th\ column density is  $0.94\times10^{15}$ \c2.
}
\ef

\begin{center}
\bf

\includegraphics[scale=0.8]{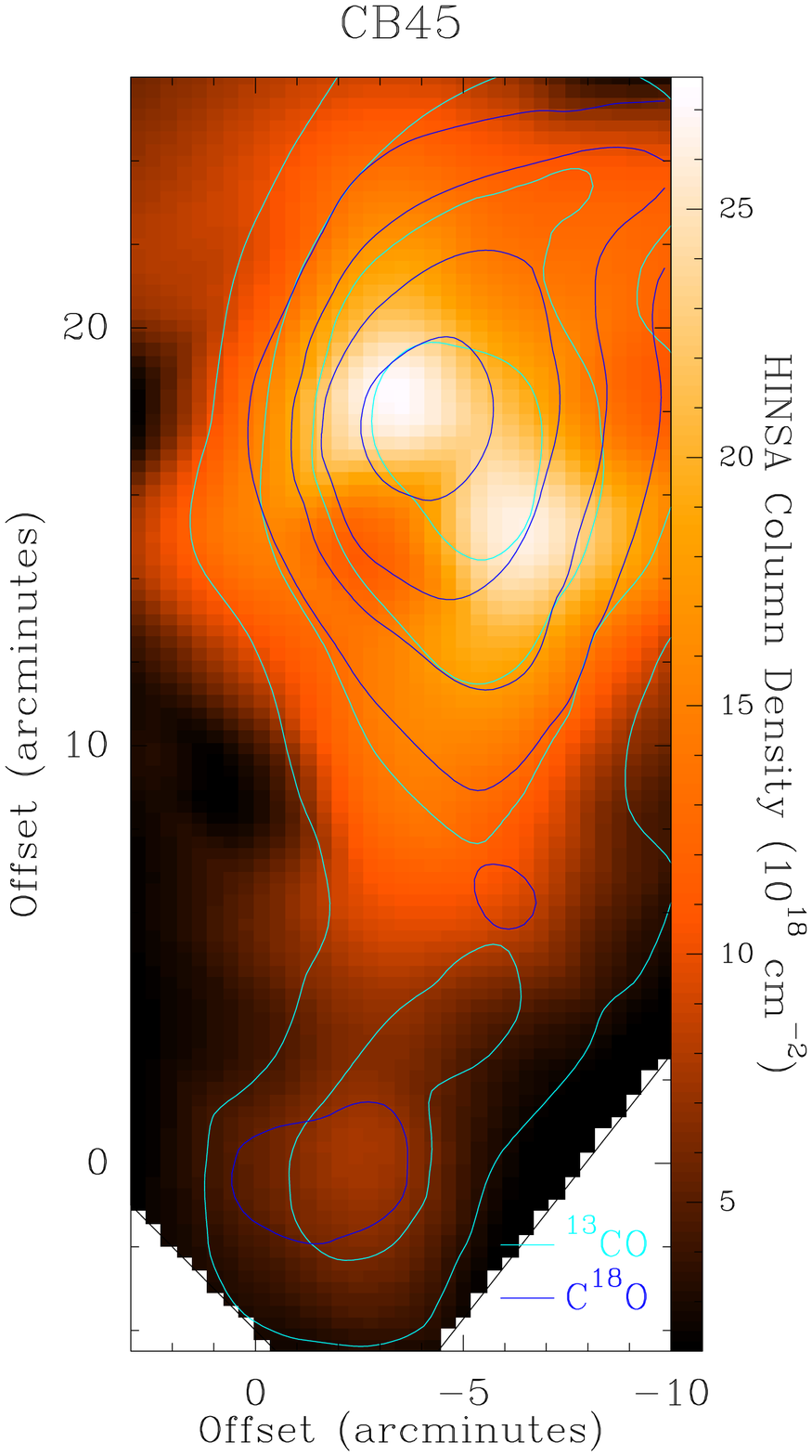}
\caption{\label{cb45} HINSA, \th\ and \ce\ column densities for CB45.
The color scale and contours are the same as in Figure \ref{l1544}.
The maximum values of the contours are $6.2\times10^{15}$  \c2\ for 
\th\ and $0.68\times10^{15}$ \c2\ for \ce.
}
\ef
\clearpage

\end{center}

\subsection{CI}
\label{CI}

The only source sufficiently extended for its distribution to be probed by
SWAS is L1544.
In Figure \ref{cil1544} we show the integrated intensity of CI and the
column densities of \th\ and cold HI (HINSA)  along a cut in Right Ascension 
through the center of L1544.  
As indicated by the two--dimensional maps presented above, 
the \th\ and the cold HI basically track each other.  
The CI, however, shows a much less centrally--peaked distribution.  
Its intensity has dropped by a factor of less than two from the center to the
edge of the map, while the \tw\ and cold HI have diminished by a factor 
$\simeq$ 5.  
While the optical depth of CI may not be negligible, this data is strongly suggestive
that the CI is predominantly in a relatively extended structure, and that the
dense core with associated large column density seen in most molecular species 
does not dominate the CI emission from this dark cloud.  
This picture is supported by the relatively large line widths of the CI.
As discussed in \citet{li2003}, these are generally much broader than those of HINSA
or the carbon monoxide isotopologues.
As shown in Figure 7 of that paper, the \th\ and HINSA line widths, which
are largely nonthermal, are well correlated, while the CI line widths are essentially
uncorrelated with those of HINSA and molecular species.
In the case of L1544 our improved data here\footnote{The line width given here 
is narrower than that given in \citet{li2003}, presumably due to  
the higher signal to noise ratio.}
give a CI FWHM line width of typically 1.5 \kms.
Comparison with the $\simeq$ 0.5 \kms\ FWHM line widths of \ce\ and \th indicates that
the more extended region in which the CI is abundant has a larger turbulent velocity
dispersion than does the region in which the molecular emission and HINSA absorption
are produced.

\clearpage
\bf
\includegraphics[scale=0.8]{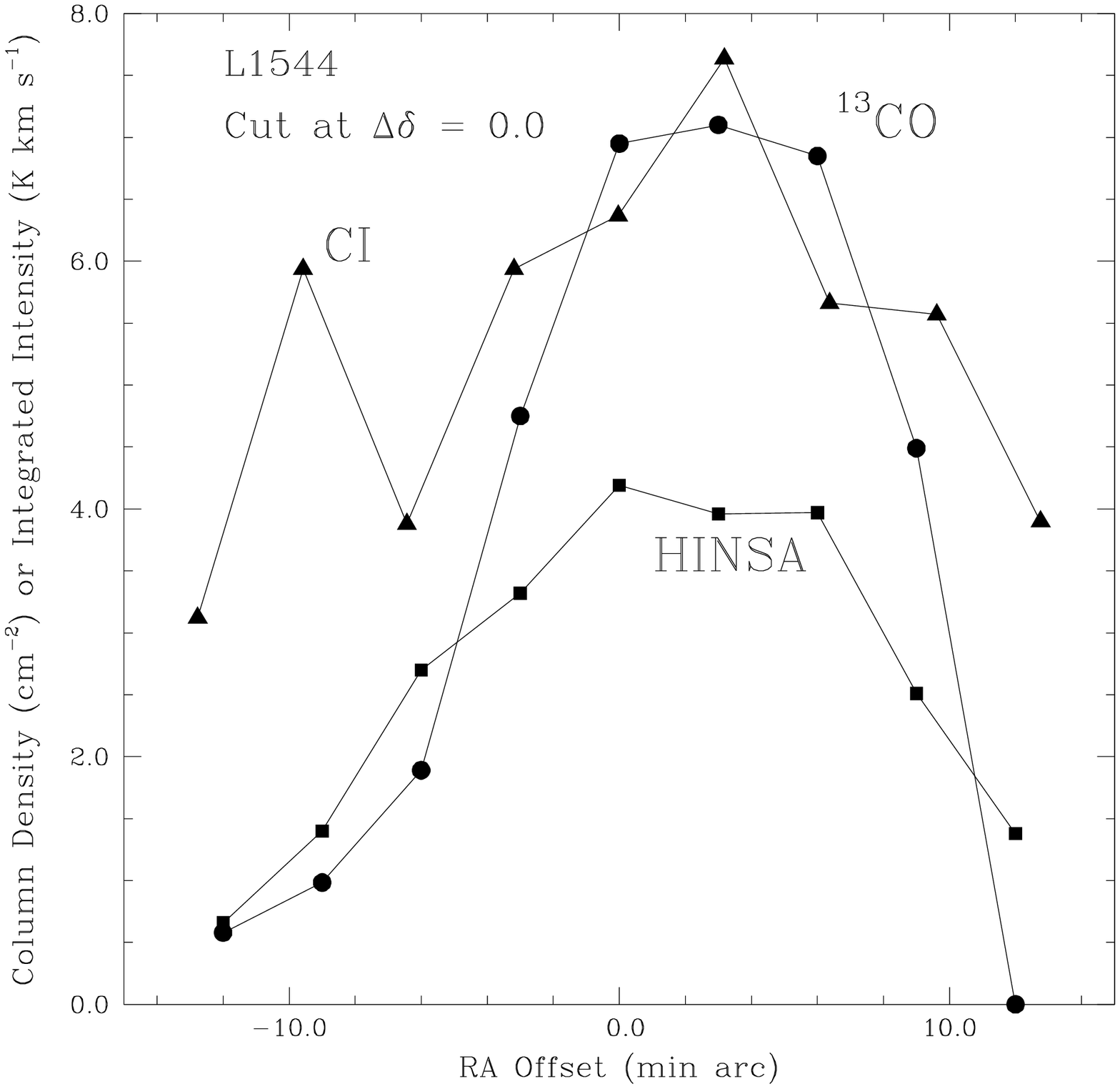}
\caption{\label{cil1544} Cut through the center of L1544 showing column density of
\th\ (divided by a factor 1$\times10^{15}$), the column density of cold HI producing
the narrow HINSA feature (divided by a factor 1$\times10^{18}$), and the integrated
intensity of the CI $^3$P$_1$ -- $^3$P$_0$ transition (multiplied by a factor of 4).
The CI emission is clearly more spatially extended than the molecular emission or the HI
absorption.
}
\ef
\clearpage

\section{CORRELATIONS OF COLUMN DENSITIES}
\label{corr}

\subsection{Carbon Monoxide Isotopologues}

In order to trace as well as possible the molecular content of these clouds,
we have compared the emission from \th\ and \ce.  
We see from
Figure \ref{1318}, where we plot N(\th)/N(\ce) as a function of the \ce\ column
density, that this ratio decreases systematically as N(\ce) increases, or
as we move from the outer to the inner portions of the clouds.
As mentioned above, the \th\ emission in the central part of L1544 suffers from
moderate saturation, and we have corrected the optical depth and column density
for this effect, as discussed above in Section \ref{colspecdist}.

\clearpage
\bf
\includegraphics[scale=0.8]{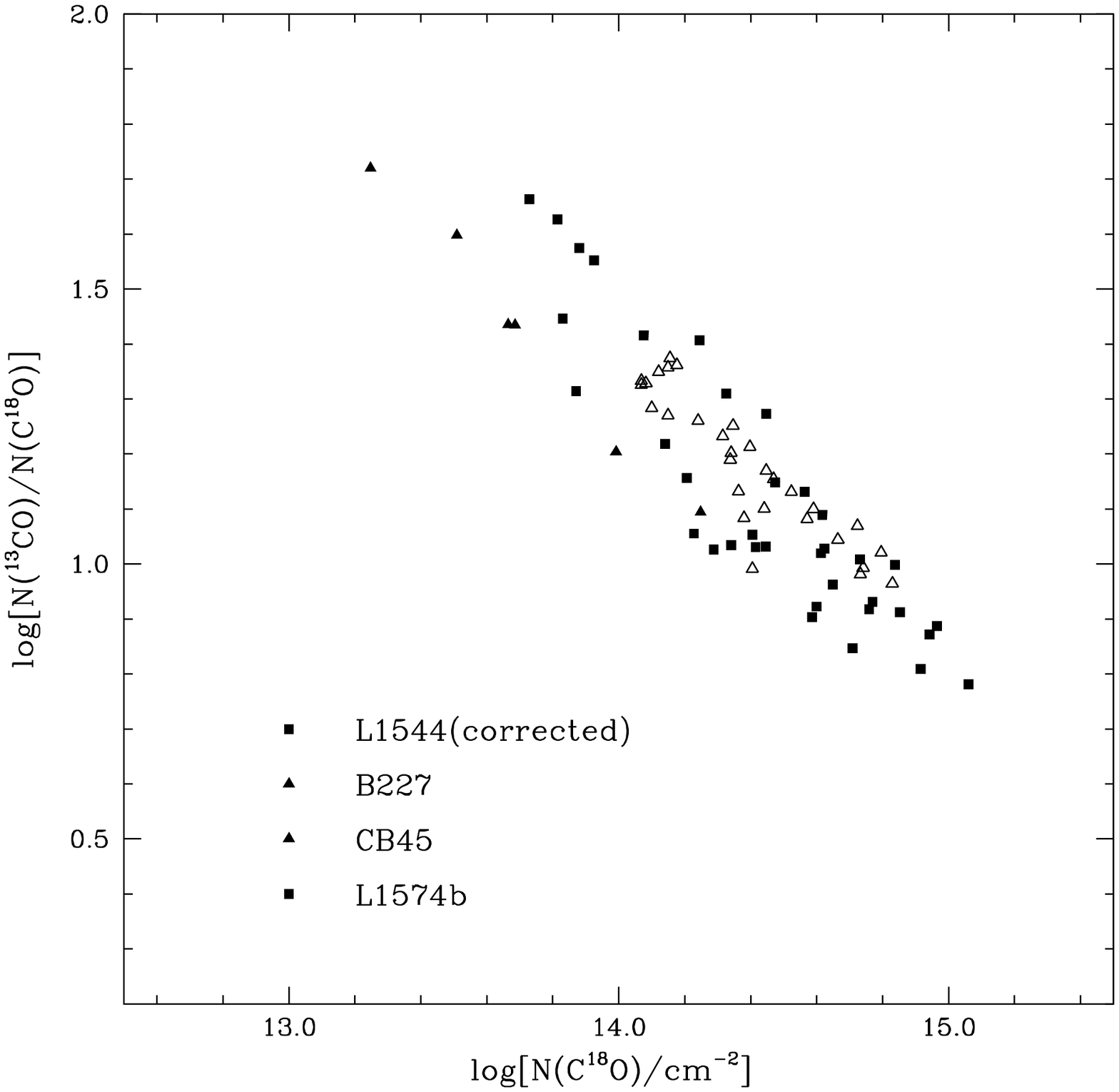}
\caption{\label{1318} Ratio of the \th\ to the \ce\ column density as a function of N(\ce).
There is no data for the red component of L1574 as \ce\ has not been reliably
measured in the 3.5 km s$^{-1}$ component of the cloud.
The points for L1544 have been corrected for saturation as discussed
in the text.
}
\ef
\clearpage

It is possible that the strong dependence of N(\th)/N(\ce) on the column density of
\ce\ is a result of uncorrected saturation in the more abundant species, but
this seems unlikely for the following reasons.  First, the L1544 data after correction
for modest saturation (a factor $\le$ 2 for the most saturated positions) agrees
well with that of the other sources.  Second, the line temperatures
at the central positions having the strongest emission in
the three other sources (Table \ref{tpeak}), are significantly below the expected 
kinetic and J = 1$\rightarrow$0 excitation temperatures.  Thus, saturation 
appears unlikely for the central positions of the three other sources, and
very unlikely for the outer parts of these clouds.  
Third, the actual values of the \th\ to \ce\ ratio at the central positions of 
L1544, CB45, and L1574b (the peak of the blue velocity component at offset (0,9) 
relative to the cloud center) are all in the range 6 -- 10, while the center of B227
has a ratio of 12.6.  These values are basically consistent with the expected oxygen
and carbon isotopic ratios \citep{langer1990, langer1993}.

Our interpretation is that the \th\ to \ce\ ratio is being increased by 
isotope--selective processes operative in the outer parts of these dark clouds.  
Processes include chemical isotopic fractionation \citep[e.g.][]{watson1976} and 
isotope--selective photodissociation \citep[e.g.][]{bally1982}.  
The chemical isotopic fractionation is particularly effective in cold regions,
but requires the presence of significant fraction of carbon in ionized form to
operate.  The isotope--selective photodissociation relies on reduced rates
for this process resulting from self--shielding in line destruction channels.
Studies including both processes indicate that chemical isotopic fractionation
will be the dominant mechanism increasing the \th\ to \tw\ ratio and the
\th\ to \ce\ ratio at visual extinctions between 0.5 and 1 mag, 
if the gas temperature is $\simeq$ 15 K \citep{chu1983, vandishoeck1988}.
Both may work to increase the \th\ to \ce\ ratio and both become more effective
in regions of reduced extinction where either the abundance of ionized carbon or
the rate of photodestruction of carbon monoxide isotopologues is greater.

Plotting the ratio as a function of N(\th) also shows a strong inverse correlation,
similar to those of \cite{langer1989} for a single cloud, B5. 
Those authors found a maximum of the \th\ to \ce\ ratio of approximately 30,
close to the peak in Figure \ref{1318}.  We do not, however, find the drop in the
column density ratio seen at the very outer edge of the cloud found by 
\cite{langer1989}.  
We feel that the results presented here suggest that
\th\ is the superior probe of column density in the outer portions of these clouds.  
This is because the fractionation enhances the abundance of
\th\, and thus compensates for the overall drop in the molecular abundances in
the outer portions of the cloud resulting from photodestruction.  
The good correlation of \th\ with visual extinction in the outer parts of clouds
is further confirmation of this fortuitous cancellation of opposing effects.

\subsection{OH}

We have obtained OH emission spectra simultaneously with those of atomic hydrogen, and 
in Figure \ref{oh} we present a comparison between the OH column density and that of \th.  
For the sources other than L1544 there is no clear correlation, and while there 
is some segregation of the various clouds in terms of their total \th\ column density, 
there is no distinction to be made among them in the column density of OH.
While a detailed explanation is beyond the scope of this paper, it does suggest
that a large fraction of the OH emission resides in a ``skin'' of material in 
each cloud, and thus once a certain minimum column density (or visual extinction)
is reached, the amount of OH present along the line of sight is essentially 
independent of the total column density.  N(OH) shows a large scatter which may 
reflect the environment, the cloud density, or some combination of effects.

L1544 is the single cloud which the column densities of the two species are
clearly correlated.
The column densities are related as $<$N(OH)$>$ = 10$^{-1.6}<$N($^{13}$CO)$>$ and 
N(OH) $\propto$ [N($^{13}$CO)]$^{0.6 - 0.7}$.
The difference in behavior between this cloud the others observed may be related to
its higher density and/or the more evolved state of this source.
\bf
\includegraphics[scale=0.8]{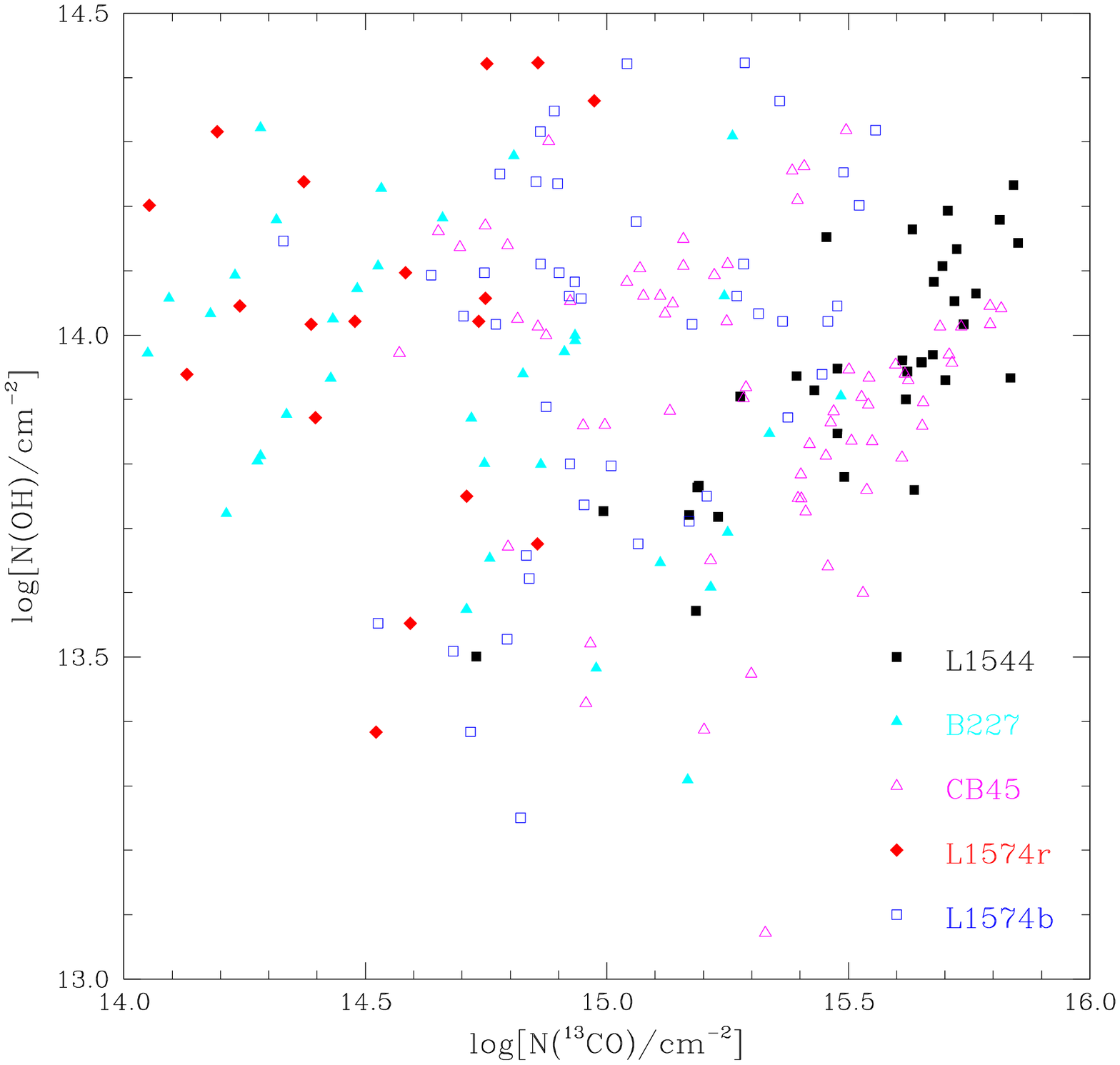}
\caption{\label{oh} Column density of OH plotted as a function of the column
density of \th.
The \th\ data for L1544 have been corrected for saturation as discussed in the text.
}
\ef

\subsection{Carbon Monoxide and Cold HI}

In Figure \ref{hinsa13co} we show the correlation of the column densities of
\th\ and cold HI in the four sources included in this study.

\clearpage
\bf
\includegraphics[scale=0.8]{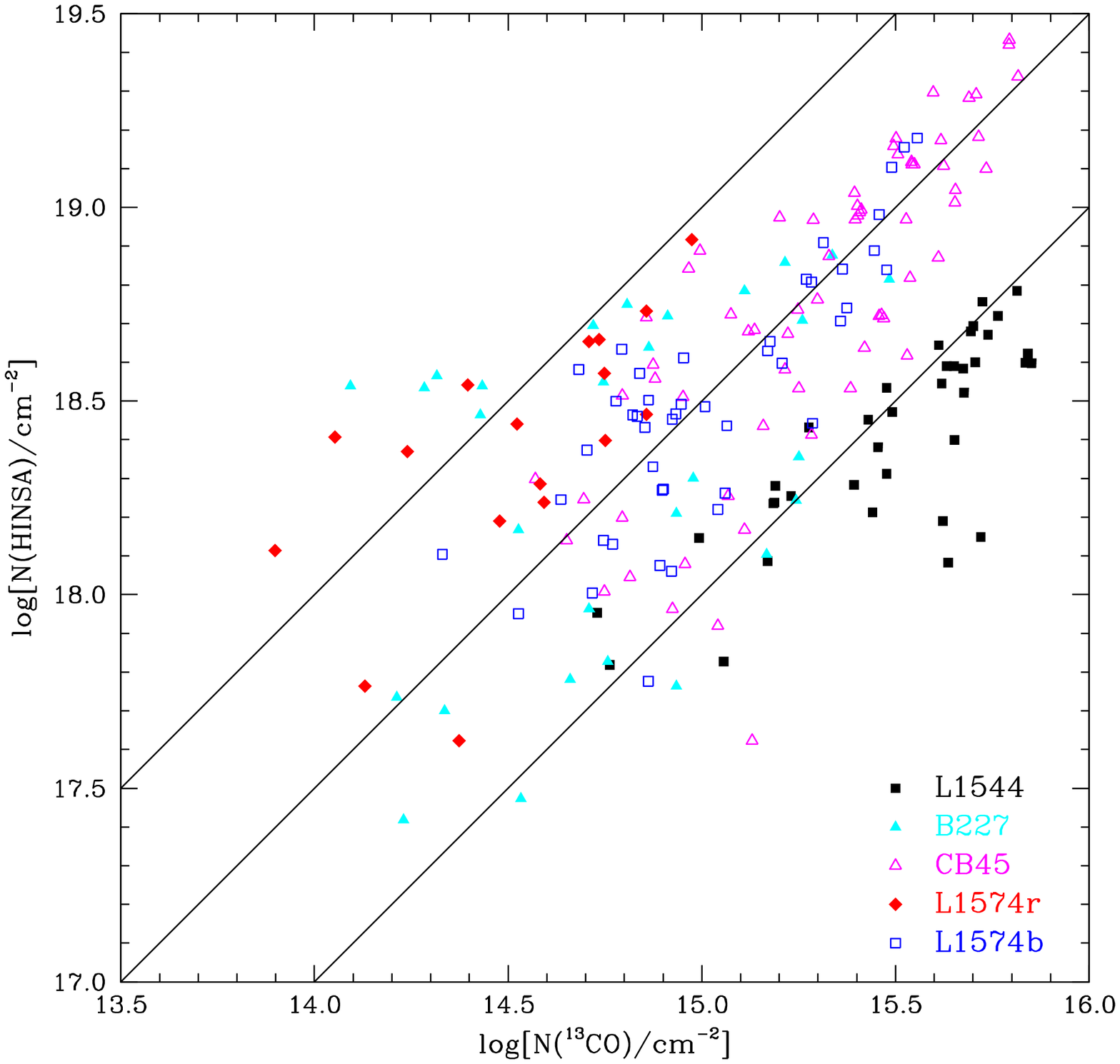}
\caption{\label{hinsa13co} Plot of the column density of cold HI (HINSA)
versus that of \th.  
The \th\ data for L1544 have been corrected for saturation.
The three diagonal lines are not fits but represent ratios N(HINSA)/N(\th) =
10$^{4.0}$ (highest), 10$^{3.5}$, and 10$^{3.0}$ (lowest).
}
\ef
\clearpage

Although there are some outlying points, it is clear that the
column density of cold HI is highly correlated with that of \th.  It is also clear
that there is some difference in the ratio of the two column densities from one
cloud to another.  L1544 has the lowest ratio, $\simeq$ 650, B227, CB45, and
L1574b are fit quite well by a ratio of 3100, and L1574r, by a ratio of 6300.
Converting the \th\ column densities to total molecular column densities
is complicated by the varying extinction that probably characterizes these
different lines of sight.  For reference, if we take a characteristic ratio
N(\th)/N(\h2) = \ 1.4$\times10^{-6}$  
\citep[the value for Taurus from][omitting the offset term]{frerking1982}, 
the highest line in Figure \ref{hinsa13co}
correponds to a ratio N(HI)/N(\h2) = 0.014, and the lowest line to a ratio a
factor of 10 smaller. 

Thus, while there is approximately a factor of 10 difference in N(HINSA)/N(\th)
between the cloud with the smallest and that with the largest value of this ratio, 
there is a clear correlation within each cloud which covers a range of $\simeq$ 20 in
the column density of either species, and for the ensemble of clouds, covering a range
of 100 in column density.  

The values for the fractional abundance of HI that we have obtained are close to
but somewhat larger than expected in a steady--state, and the implications
of this are discussed in Section \ref{discussion}.

\section{ATOMIC AND MOLECULAR DENSITIES}
\label{atmoldensities}

In order to extract the maximum value from the data about atomic hydrogen in
molecular clouds, it is helpful to determine the actual number densities of
different species.  To do this, we model the density of HI, \ce, and \th\ in
each cloud as a triaxial Gaussian, characterized by a peak value and three widths.
We determine the two widths on the plane of the sky from the extent of the
emission between half power points along two perpendicular axes.  
These are taken along (albeit imperfect) axes of symmetry for L1544, and 
along the directions of Right Ascension and Declination for the other three clouds. 

The use of a Gaussian is reasonable given the column density distributions
and the moderate dynamic range of our data. 
Given the $\simeq$ 3\arcmin\ angular resolution, we do not see any strong
central peaking characteristic of the power law distributions often found for cloud cores;
the small, dense regions are presumably significantly beam diluted in the present study. 
A Gaussian function makes it relatively easy to model the significant asymmetry seen
to be characteristic of our sources, which is difficult to do with a power law density
distribution.
Using the distances given in Table \ref{sources}, we obtain the FWHM source dimensions
given in Table \ref{densities}.
The FWHM along the line of sight is taken to be the geometric mean of the two
measured dimensions.
The central density of species X is then given by 
\be
n_{cen}(X) = \frac{0.94 N_{cen}(X)}{\Delta z_{FWHM~LOS}(X)}\lp
\ee

The HINSA and \th\ sizes for L1544 are lower limits due to 
incomplete coverage of the emission.  
This cloud is immersed in a region of lower column density \citep{snell1981} 
which extends to the northwest.  
The more extended maps suggest that the half power dimensions in this
direction are only slightly greater than our limits.
We see that the cloud dimensions in HINSA and \th\ emission are comparable, which suggests that
these two species are coextensive.  
This is consistent with the fact that the \th\ nonthermal line widths are relatively close to 
those of HINSA \citep{li2003}.  

The regions of \ce\ emission are smaller than those of HINSA or \th, but by a factor less than two.
The \ce\ line widths are correspondingly smaller than those of HINSA.
Thus, while the two molecular tracers both delimit regions comparable in size to that of
the cold HI, it appears that the \th\ is probably the better tracer for comparison with the atomic
hydrogen to determine fractional abundances.
The central densities given in column 5 of Table \ref{densities} are intended to be representative
of the central portion of the cloud, but given the moderate angular resolution of our
observations and the fairly gradual fall--off of the densities, it is characteristic of 
a significant fraction of the cloud volume.

In order to determine the \h2\ densities in the centers of the cloud, we
have to adopt fractional abundances for \ce\ and \th; these inevitably
are somewhat uncertain.  
In particular, depletion of carbon monoxide isotopologues onto dust grains,
while not so evident as for some other species, is a potential problem for
the densest, most quiescent regions of dark cloud cores \citep[e.g.][]
{caselli1999, bacmann2002, bergin2002, tafalla2002, redman2002}.
However, all of these studies indicate that significant depletion of 
rare isotopologues of carbon monoxide is restricted to the central,
densest regions of clouds.  

The map of CCS emission in L1544 \citep{ohashi1999} shows an elliptically shaped
region approximately 1\arcmin.5 by 3\arcmin\ in extent.
There is evidently a central region in which is molecule is significantly
depleted, but it is between 1\arcmin\ and 1\arcmin.5 in diameter.
A similar--sized region is heavily depleted in L1498 \citep{willacy1998}.
For more distant clouds, regions of similar physical size 
(1.3 -- 3 $\times$10$^{17}$ cm) would appear smaller.
These ``depletion zones''  are heavily beam diluted (by a factor $\simeq$ 10 in the single
beam  looking at the center of the cloud), and would of course be irrelevant
for the vast majority of positions we have studied.

The region over which depletion has a major impact can also be seen in terms 
of the density required for significant depletion, which is  
$10^5$ -- $10^6$ cm$^{-3}$ \citep{bacmann2002, tafalla2002}.
This is basically two orders of magnitude greater than the average densities
we find, consistent with the depletion being confined to a small portion
of the line of sight as well as of the 3\arcmin.3 beam.
Using empirically determined density profiles, \cite{tafalla2002}
find that characteristic radii within which this density is attained are
$\simeq$ 10$^{17}$ cm, or 1\arcmin.

The depletion rate is the rate of collisions which result on sticking to a grain.
For molecules, the sticking coefficient $S$ is taken to be unity \citep{burke1983},
while for HI (see Section \ref{reh2form} below), $S$ = 0.3. 
The depletion rate is proportional to the sticking coefficient and inversely
proportional to the square root of the particle's mass \citep[e.g.][]{draine1985}.
The depletion time scale is the inverse of this rate, which gives 
\be
\frac{\tau_{CO~depletion}}{\tau_{HI~depletion}} = 1.6\lp
\ee
Scaling the results of Section \ref{reh2form} for molecular hydrogen formation 
to atomic hydrogen depletion and then to carbon monoxide depletion, 
we obtain $\tau_{CO~depletion}$ = 2.1$\times10^9$/n$_{H_2}$ yr.
This is consistent with the results of \cite{caselli1999} and \cite{bacmann2002}, that
$\tau_{CO~depletion}$ $\simeq$ 10$^4$ yr for n$_{H_2}$ = 10$^5$ cm$^{-3}$.
The relatively short time scale for molecular depletion is thus a result of
the very high densities in the cloud cores which overall have shorter
evolutionary time scales than those characteristic of the lower density clouds 
in which they are embedded.

The modest effect of the depletion on determining the column density even
in the beam including the cloud core is illustrated by the result of
\cite{caselli1999} who derive that the ``missing mass'' of H$_2$ from
depletion of tracers is 2.3 solar masses in L1544.  
The column density in the central 3\arcmin.3 beam assuming undepleted
abundance of \ce\ is 4 solar masses.
Thus, we could be underestimating the H$_2$ column density by as much as 50\% in
this direction.
This would have the effect of decreasing the HI to H$_2$ abundance
ratio in this (and possibly other) core directions, but would not
change to an appreciable degree the correlation of HI and H$_2$ or
the implications of the ratios determined.

Using the molecular hydrogen column densities and the source sizes, 
we derive the central densities of atomic and molecular 
hydrogen given in Table \ref{atmoldens}.
We see that the atomic hydrogen densities are between 2 and 6 cm$^{-3}$. 
The density of molecular hydrogen derived from \ce\ is on the average a
factor 1.7 larger than that derived from \th.  
As discussed above, the \th\ is probably a better tracer of the large--scale
distribution of gas, but is more likely to miss material in the central
region of the cloud due to uncorrected saturation.  
The results from both tracers are in reasonable agreement and we should
have results from both in mind for comparison of the fraction of atomic
gas measured here with that predicted from models presented in the
following section.

\section{TIME--DEPENDENT MODELING OF THE ATOMIC TO MOLECULAR HYDROGEN RATIO}
\label{modeling}

\subsection{Rate Equation for \h2\ Formation}
\label{reh2form}

In order to calculate the gas phase abundances of atomic and molecular
hydrogen, we need to model the grain surface production \h2.
For the present study, we assume that all dust grains are
spherical particles, differing only in their size.  
This neglects the changes in grain density and structure that
could be a consequence of processes which initially formed the dust
grains, or modified them by accretion or coagulation.  
But since the effects of grain evolution, particularly on the surface
properties of grains which are critical for \h2\ formation and 
molecular depletion are so uncertain, this simplification seems a 
reasonable starting point.

The basic equation giving the formation rate of \h2\ molecules on
the surfaces of grains of a single type, e.g. grains having a
specified cross section and number density \citep{hws71}, is
\be
\label{rh2single}
R_{H_2} = \frac{1}{2} S_{HI} \epsilon_{H_2} n_{HI}<v_{HI}>\sigma_{gr}n_{gr}~.
\ee
$R_{H_2}$ has units of cm$^{-3}$ s$^{-1}$, $S_{HI}$ is the probability
of a hydrogen atom which hits a grain sticking to it, and $\epsilon_{H_2}$
is the probability of a hydrogen molecule being formed by recombination of 
two hydrogen atoms and then being desorbed from the grain\footnote{
$S_{HI}$ and $\epsilon_{H_2}$ were multiplied together to form the \h2\ 
recombination coefficient $\gamma$ by \citet{hws71}, but following more recent usage
we separate the two terms here.  There remains some inconsistency in
the terminology used, as e.g. \citet{cazaux2002} use the expression 
``recombination efficiency'' for $\epsilon_{H_2}$, which clearly does not
include the sticking probability $S_{HI}$.
}.  
The atomic hydrogen density
in the gas is $n_{HI}$ (cm$^{-3}$), the mean velocity of these atoms is 
$<v_{HI}>$ (cm~s$^{-1}$), while the number density of grains having 
cross section $\sigma_{gr}$ (cm$^{2}$) is $n_{gr}$ (cm$^{-3}$).

The sticking coefficient, estimated to be 0.3 by \citet{hollenbach71}, has
been studied in a quantum mechanical treatment by \citet{ldw85}.  
These last authors considered physisorption and chemisorption on
plausible grain surfaces of graphite, silicate, and graphite covered with
a monolayer of water.  
The values of $S_{HI}$ found range from 0.9 to very small values.
Given the uncertainties in grain composition and binding mechanisms, 
it appears reasonable to adopt the value $S_{HI}$ = 0.3.

For grain surfaces with binding sites of a single type, there is an
inherent competition between thermal desorption and \h2\ formation.
If the binding is very strong, atoms on the surface will not desorb,
but they will not be mobile, so the \h2\ formation rate will be low.
On the other hand, if the binding is very weak, the mobility will
be high, but so will be the rate of thermal desorption, and again, the
rate of formation of \h2\ will be reduced.
This has led to the concern that the formation rate of molecular
hydrogen will be large only over a very narrow range of grain temperature,
a result which would certainly be problematic given the wide range
of temperatures characterizing clouds in which significant fractions
of molecular hydrogen are found.

\citet{cazaux2002} circumvented this problem by postulating two types
of binding sites on individual grains - physisorption sites with
relatively weak binding, and chemisorption sites characterized by 
10 to 100 times stronger binding.  
At the temperatures of dense molecular clouds, there will always
be atoms in chemisorption sites, because the thermal desorption
time scale far exceeds any other.  
Thus, any hydrogen atom hitting the grain and sticking will end up
in a physisorption site, from which it can tunnel to a chemisorption
site where it finds a partner.  
The barrier between the two types of sites is moderate (E/k = 200 K),
so that the effective mobility is high\footnote{
This is consistent with the most comprehensive treatment, which suggests that
hydrogen atoms will be highly mobile under any reasonable conditions
\citep{ldw84}.
}.  
\citet{cazaux2002} show that for grain
temperatures between 6 K and 30 K, $\epsilon_{H_2}$
is unity, and drops gradually to 0.2 for a grain temperature of 100 K.
We adopt here $\epsilon_{H_2}$ = 1.0.

The grain mass per unit volume implied by
the single type of grain assumed in equation \ref{rh2single} is given by
\begin{equation}
\label{gdr}
n_{gr}m_{gr} = \frac{n_{gas}<m_{gas}>}{GDR}~,
\ee
where $m_{gr}$ is the mass of a grain, $n_{gas}~(\simeq n_{H_2} + n_{HI} + n_{He})$  
is the total volume density of gas particles, 
$<m_{gas}>$ is the average mass of a particle in the gas, and  
$GDR$ is the gas to dust ratio by mass.
The average mass per gas particle is
\begin{equation}
\label{mgasavg}
<m_{gas}> = \frac{(2n_{H_2} + n_{HI} + 4n_{He})m_p}{n_{gas}}~,
\ee
where $m_p$ is the mass of the proton.
With this, we can write
\begin{equation}
\label{nsigma}
n_{gr}\sigma_{gr} = n_{gas}\sigma_{gr}\frac{<m_{gas}>}{m_{gr}GDR}~.
\ee

The mass of a grain can be written $m_{gr} = \rho_{gr}V_{gr}$,
where $\rho_{gr}$ is the density of a grain and $V_{gr}$ is its volume.
Since we have no clear evidence that the grain density varies systematically
with grain size, we adopt a single value of the grain density
(typically $\simeq$ 2 gm~cm$^{-3}$) for grains of all sizes.  
Associating $\rho_{gr}$ with the other grain parameters assumed to be
independent of grain size, we can rewrite the H$_2$ formation rate as
\be
\label{rh2-2}
R_{H_2} = \frac{S_{HI}\epsilon_{H_2} <v_{HI}> <m_{gas}>}{2\rho_{gr}~GDR}
		\frac{\sigma_{gr}}{V_{gr}} n_{gas} n_{HI}~.
\ee

For grains of a single size, the \h2\ formation rate is proportional to
the ratio of the grain cross section to its volume. 
We can write the \h2\ formation rate as
\be
	\label{avgRH2}
	R_{H_2} = k_{H_2} n_{gas} n_{HI}\lc
\ee
where all of the other factors in equation \ref{rh2-2} have been subsumed into 
\be
	\label{KH2}
	k_{H_2} = \frac{S_{HI}\epsilon_{H_2}<v_{HI}><m_{gas}>}{2\rho_{gr}GDR} 
	\frac{\sigma_{gr}}{V_{gr}}\lp
\ee
For a single ``standard'' grain species of radius $a_s$, 
$\sigma_{gr}/V_{gr}$ = $3/4a_s$.

In Appendix A, we evaluate the effect of a grain size distribution
on the rate coefficient for \h2\ formation.  
With the simplest assumptions that all grains have the same density,
and that the sticking coefficient and formation efficiency are independent
of grain size, the effect of a power law distribution of grain
radii can be analyzed straightforwardly, and for the 
Mathis, Rumpl, \& Nordsieck (1977; hereafter MRN) 
size distribution, we find the particularly simple result given
by equation \ref{mrn-unnorm} that $k_{H_2}$ depends only on the maximum
and minimum values of the grain size distribution.  

Defining the total proton density in atomic and molecular hydrogen as
\be
	\label{totprot}
	n_0 = n_{HI} + 2n_{H_2}\lc
\ee
and including a standard helium abundance of 0.098 relative to hydrogen in
all forms, we can write
\be
	n_{gas}<m_{gas}> = (1 + \frac{4n_{He}}{n_0})m_pn_0 = 1.39m_pn_0 \lc
\ee
which gives
\be
	\label{rh2-3}
	R_{H_2} = k'_{H_2~MRN}\nh1n_0\lc
\ee
with 
\be
k'_{H_2~MRN} = \frac{S_{HI}\epsilon_{H_2}<v_{HI}>1.39m_p}{2\rho_{gr}GDR}
\frac{3}{4a_s} \frac{a_s}{\sqrt{a_{max}a_{min}}} \lp
\ee
The final fraction represents the correction due to the grain size
distribution, being equal to unity for any single grain size.
Substituting standard values for the relevant parameters, 
and normalizing to them, we find
	\footnote{\label{hiturb}Following almost all treatments of \h2\ formation, we include
	only the thermal velocity of the hydrogen atoms in equation 
	In principle, we might also consider the nonthermal component of the line widths, 
	which is generally considered to be turbulent in nature. 
	However, this will affect processes such as the collsion of a hydrogen atom
	with a grain only if the characteristic size scale of the turbulence is
	comparable to or smaller than the mean free path of the atoms.  Taking the most
	conservative approach including	only collisions between atoms and grains, this
	is approximately 10$^{16}$/n$_{gas}$ cm.  This very small distance is far smaller
	than the scale size on which line widths in cloud cores approach their 
	purely thermal values \citep{barranco1998, goodman1998}.  Thus it appears
	justified to include only the thermal velocity in equation \ref{formratenorm}.,
	which gives the \h2\ formation rate varying as $T^{0.5}$.
	Fortunately, the nonthermal contribution to the linewidth for HINSA is,
	unlike the situation for HI emission and most absorption, quite modest
	compared to the thermal linewidth (as seen in Table 2 of \citet{li2003} and Table 6 
	here) and thus the uncertainty introduced by considering only the
	thermal velocity for the formation rate is less than a factor of 2.}
\be
\label{formratenorm}
	k'_{H_2} = 3.5\times10^{-18}
  	\frac{(\frac{S_{HI}}{0.3})(\frac{\epsilon_{H_2}}{1.0})({\frac{T}{10 K}})^{0.5}}
  	{(\frac{\rho_{gr}}{2 g cm^{-3}})(\frac{GDR}{100})(\frac{a_s}{1.7\times10^{-5} cm})}
  	\frac{a_s}{\sqrt{a_{max}a_{min}}}~cm^3s^{-1}
  	\lp
\ee

Based on the discussion in Appendix A, and the values for a \citet{MRN}
grain size distribution with $a_{min}$ = 25 \AA, $a_{max}$ = 10,000 \AA,
and $a_s$ = 1,700 \AA, we find that the effect of the grain size distribution is to
increase the formation rate coefficient by a factor of 3.4.
We thus take a nominal value for $k'_{H_2}$ = 1.2$\times10^{-17}$ cm$^3$s$^{-1}$.

The numerous factors which enter into the \h2\ formation rate coefficient
inevitably result in a relatively large uncertainty in this parameter.
The range of  values for the sticking coefficient $S_{HI}$  and the
H$_2$ formation efficiency $\epsilon_{H_2}$ for ``traditional'' grains
have been discussed earlier in this section.  
The grain density for grains of standard composition can vary by a factor 
$\simeq$ 1.5, and a similar variation in the gas to dust ratio is entirely
plausible, although data in dark clouds is very sparse.  
The upper and lower limits on the grain size distribution enter as
$(a_{max}a_{min})^{-0.5}$, but these quantities could easily differ from
our adopted values by a factor $\simeq$ 2.
It is also not clear whether very small grains would be effective for producing
molecular hydrogen as their surface properties are likely very different from
grains studied theoretically or measured in the laboratory.  
Combining the uncertainties in the various factors, 
$k'_{H_2}$ may differ from its nominal value by a factor of 5.
This directly impacts the \h2\ formation time scale and the steady--state
HI density, both of which vary as  ${k'_{H_2}}^{-1}$.

\subsection{Time Dependence of HI and \h2\ Densities}
\label{timedep_sec}

An idealized scenario for the evolution of the abundance of HI and \h2\
consists of a cloud in which the hydrogen is initially entirely in atomic form, 
maintained in this state by photodestruction.
At t = 0 the extinction is suddenly increased so that photodestruction in the bulk
of the cloud volume (regions with visual extinction greater than $\simeq$ 0.5 mag) can
be neglected. 
This is justified since the photodissociation rate of \h2\ drops dramatically 
as a function of increasing column from the cloud surface, due to effective 
self--shielding \citep[e.g.][]{vandishoeck1988, draine1996}.

To model the evolution of this cloud, 
we assume that the only destruction pathway for molecular hydrogen is
cosmic ray ionization 
	\footnote{
	There are cases in the literature where it is unclear whether one is dealing
	with the cosmic ray ionization rate per H atom or the rate per H$_2$ molecule 
	(in molecular clouds).
	Since the regions studied here are largely molecular, we adopt the latter 
	definition.}
, which is assumed to occur at a rate $\zeta_{H_2}$ s$^{-1}$.  
The value of $\zeta_{H_2}$ in dense clouds has been found to lie in the range
$10^{-18}$ to $10^{-16}$ s$^{-1}$ \citep{caselli1998}, but in other
studies has been more narrowly defined to be equal to $5\times10^{-17}$ s$^{-1}$
\citep{bergin1999}, and $(5.2\pm3.6)\times10^{-17}$ s$^{-1}$ \citep{vandertak2000}.  
This is consistent with results of \citet{doty2002} from detailed modeling of
the source AFGL2591.

A much higher ionization rate, $\zeta_{H_2}$ = $1.2\times10^{-15}$ s$^{-1}$  
found in diffuse clouds \citep{mccall2003}, can possibly be reconciled with the above if
there is a component of the cosmic ray spectrum which penetrates diffuse
but not dense clouds.  
\citet{doty2004} suggest that the much higher rate which they derive
for IRAS 16293-2422 is actually being dominated by X--ray ionization
from the central star, and that the true cosmic ray ionization rate is
much lower.
We here adopt the average of the values obtained by
van der Tak \& van Dishoeck, $\zeta_{H_2}$ = 5.2$\times10^{-17}$ s$^{-1}$.
  
We assume that the evolution of the cloud takes place at a constant
density, and write the equation expressing the time dependence of the 
molecular hydrogen (denoting the formation rate coefficient
simply by $k'$) as

\be
	\label{timedep}
	\frac{dn_{H_2}}{dt} = k' n_{HI} n_0 - \zeta_{H_2} n_{H_2}\lp
\ee
Defining the fractional abundance
$x_i$ (of atomic and molecular hydrogen) as the density of species $i$ divided
by the total proton density, we find that
\be
	\frac{dx_2}{dt} = k' x_1 n_0 - \zeta_{H_2} x_2\lp
\ee
Substituting equation \ref{totprot} we can rewrite this as
\be
	\frac{dx_2}{dt} = k' n_0 - (2k' n_0 + \zeta_{H_2}) x_2\lp
\ee

The solution is the time dependence of the fractional abundance of 
molecular hydrogen, given by

\be
	\label{x2t}
	x_2(t) = \frac{k' n_0}{2k' n_0 + \zeta_{H_2}}[1 - exp(-t/\tau_{HI \rightarrow H_2})]\lp
\ee
The fractional abundance of atomic hydrogen is 1 -- 2$x_2$, and hence is
given by
\be
	\label{x1t}
	x_1(t) = 1 - \frac{2 k' n_0}{2k' n_0 + \zeta_{H_2}}[1 - exp(-t/\tau_{HI\rightarrow H_2})]\lp
\ee
The time constant for HI to \h2\ conversion is given by
\be
	\label{timeconst}
	\tau_{HI\rightarrow H_2} = \frac{1}{2k'n_0 + \zeta_{H_2}}\lp
\ee
The steady--state molecular and atomic gas fractions can be
found from equations \ref{x2t} and \ref{x1t}, but in the
limit of present interest $k' n_0$ $\gg$ $\zeta_{H_2}$ we obtain
\be
	\label{tausimple}
	\tau_{HI\rightarrow H_2} = \frac{1}{2k'n_0}\lp
\ee
The steady--state values for the constituents of the gas are then
\be
	x_2 \rightarrow \frac{1}{2}~ {\rm or}~ n_{H_2} \rightarrow \frac{n_0}{2}\lc
\ee
and
\be
	\label{at_longtime_h2dom}
	x_1 \rightarrow \frac{\zeta_{H_2}}{2 k' n_0}~ {\rm or}~ n_{HI} \rightarrow \frac{\zeta_{H_2}}{2 k'}\lp
\ee

This limit thus corresponds to our having essentially molecular gas, 
but the implication of equation \ref{at_longtime_h2dom} is that there is 
a constant density of atomic gas given by

\be
	\label{resid}
	 n_{HI}(t \gg \tau_{HI\rightarrow H_2}) \equiv n_{HI}^* = \frac{\zeta_{H_2}}{2 k'}~.
\ee

In Figure \ref{timedeph1} we show the evolution of the atomic hydrogen fractional
abundance as a function of time, for clouds in which the total proton density
is fixed.
From the discussion in Section 6.1 and Appendix A, we have adopted
$k'$ = $1.2\times10^{-17}$ cm$^3$s$^{-1}$ and $\zeta_{H_2}$ = $5.2\times10^{-17}$ s$^{-1}$.
This yields an equilibrium atomic hydrogen density $n_{HI}^*$ equal to 2.2 cm$^{-3}$.
Equation \ref{resid} is not at all new; it was given by \citet{solomon1971}, but
their use of a much higher value of $\zeta_{H_2}$ resulted in an estimate of the
equilibrium abundance of atomic hydrogen equal to 50 cm$^{-3}$, a value clearly
ruled out by our data.
\bf
\includegraphics[scale=0.8]{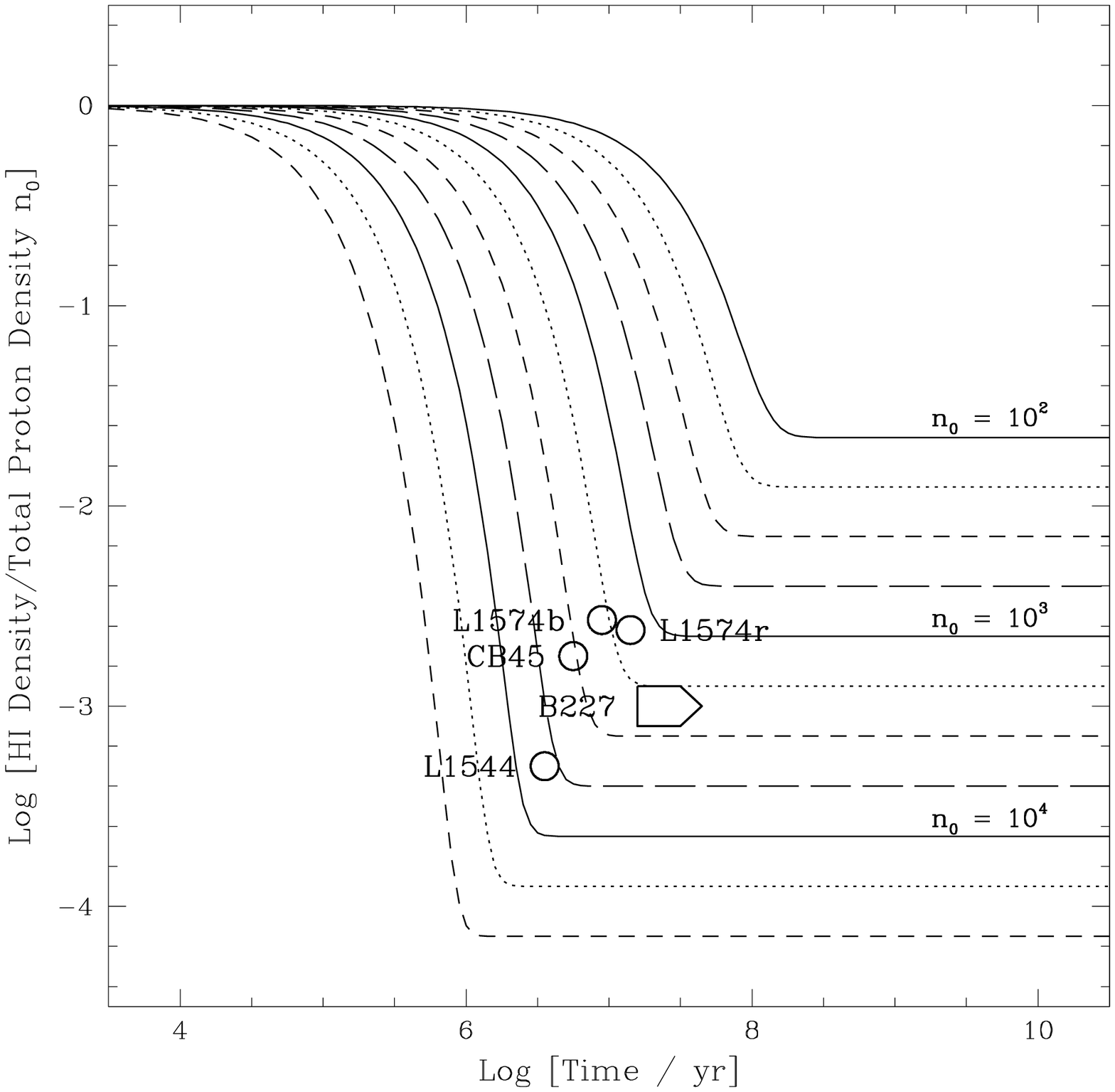}
\caption{\label{timedeph1} Time dependence of the fractional abundance of atomic
hydrogen for clouds of fixed total proton density (in cm$^{-3}$). 
The curves of different total proton densities are logarithmically 
spaced at intervals of 0.25.
The results for five sources studied here, located in terms of the atomic
hydrogen to total proton density ratio and total proton density, are also included.
The four sources with satisfactory solutions are indicated by open circles.
B227, for which the fractional HI abundance is very close to the steady
state value, is indicated by an arrowed box at the youngest age reasonably consistent
with the data.
}
\ef

\section{DISCUSSION AND IMPLICATIONS FOR CLOUD EVOLUTION}
\label{discussion}

\subsection{Cloud Age Based on Time--Dependent HI Fraction}

For the nominal values we have adopted, the simple form of $\tau_{HI\rightarrow H_2}$ given by equation 
\ref{tausimple} will apply for n$_0$ $\gg$ $n^*$ = 2.2 cm$^{-3}$.  
In this limit, which is certainly relevant for dense clouds, the time constant is
\be
	\label{taunumeric}
	\tau_{HI\rightarrow H_2} = \frac{8.3\times10^{16}}{n_0}\ s = \frac{2.6\times10^9}{n_0}\ yr\lp
\ee
Assuming that the clouds we have studied are characterized by densities determined 
from the \th\ observations, n$_0$ = 2 to 6 $\times10^3$ cm$^{-3}$, and
$\tau_{HI\rightarrow H_2}$ = 0.5 to 1.3 $\times10^6$ yr.
Using the larger densities from \ce\ reduces this 
characteristic timescale by a factor $\simeq$ two
	\footnote{
	The central densities derived from our observations do
	not reflect the true central density of the cloud because 
	the \ce~ J = 1 $\rightarrow$ 0 
	transition is not sensitive to densities $\ge$ 10$^4$ cm$^{-3}$, 
	and we have smoothed the FCRAO data to 3\arcmin.3 beamwidth for 
	comparison with the Arecibo HI data.}
.  As seen in Figure \ref{timedeph1}, the fact that the clouds observed have HI fractional abundances relatively close to
steady--state values immediately indicates that the time elapsed since 
initiation of HI to \h2\ conversion must be several times greater than $\tau_{HI\rightarrow H_2}$ or
at least a few million yr.
This ``age'' applies to the bulk of the molecular cloud as traced in \th\ and
HINSA
	\footnote{
	Again, this does not apply to the cloud cores, particularly that of L1544
	where densities are 10$^5$ -- 10$^6$ cm$^{-3}$ \citep{caselli1999,tafalla2002}
	and where the HI $\rightarrow$ H$_2$ conversion timescale
	will be much shorter.}
.  For four of the five clouds studied, we obtain a unique solution for the 
elapsed time.  
This quantity, which we denote $\tau_{cloud}$, is defined by the total
proton density (twice the molecular hydrogen density obtained from the \th\ column density
and source size), and the fractional abundance of atomic hydrogen.  
While keeping in mind the uncertainty of the ``age'' as determined here,
for these four dark clouds, this quantity lies in the range $10^{6.5}$ yr $\leq$ 
$\tau_{cloud}$ $\leq$ $10^7$ yr.  
The fractional abundance of atomic hydrogen in B227 is almost exactly the steady state
value, and choosing a minimum time when this quantity will have dropped to close to
this value leads to $\tau_{cloud}$ $\ge$ $10^7$ yr.

These relatively great cloud ages would pose a problem if there were clear evidence 
that these clouds were dynamically evolving.  
To evaluate this possibility, we have used the gaussian density distribution and the parameters
derived above.  Using the results from Appendix \ref{gaussvirial} and the masses 
and line widths given in Table \ref{virialparam}, we calculate the potential and
kinetic energies, ignoring surface effects, the external medium, and any contribution
of the magnetic field.  
The clear result is that these clouds are close to virial equilibrium, with
the possible exception of L1544, for which the numbers are particularly uncertain
due to the incompletely defined extent of the cloud, but which appears to have $-\U$
noticeably larger than $2\T$. 
While this does not define the evolutionary path these clouds have followed,
it does suggest that it is entirely plausible that they have spent several
million years evolving to their present state\footnote{
	\citet{caselli1999} suggest that the core of L1544
	is collapsing with a much shorter characteristic time scale.  
	Only a very small fraction of the cloud is involved in the collapse,
	suggesting that the evolution of this region is, in fact, decoupled from
	that of the bulk of the cloud which we are studying here.}
.  

In fact, we feel that these values for cloud ages must, for a number of reasons, be 
considered as lower limits.
First, it is unlikely that the evolution of these clouds proceeds at constant density.
Typical models of interstellar cloud evolution indicate that the path followed starts with
lower--density atomic clouds and proceeds to higher density.  
This is not a requirement, in that if, for example, a compressive event initiates the
process by compressing the gas and increasing the visual extinction, the density
could immediately be increased to something like its final value, but in fact, no
atomic clouds with even the moderate densities given in Table \ref{atmoldens} 
have been observed.  
If cloud evolution includes an increase in the density occurring simultaneously with
the HI to \h2\ conversion, then much of the evolution will have taken
place at lower densities than those now observed, and the time scale will be 
correspondingly longer. 
However, if the clouds were much more turbulent in the past and
the turbulence were on a scale that affected the velocities of hydrogen atoms
colliding with grains (see footnote \ref{hiturb}), then the lower collision 
rate and \h2\ formation rate resulting from the lower density would be 
partially offset by the higher HI--grain velocities.  

Second, we may consider what happens if the column density is not sufficient to make
photodestruction of \h2\ insignificant in the central regions of these clouds.  
To first order, we can consider photodestruction of \h2\ simply as adding to the
cosmic ray destruction rate in equation \ref{timedep}.  
While formally increasing the time constant (equation \ref{timeconst}), for clouds
of high density, $\tau_{HI\rightarrow H_2}$ is unaffected, as long as n$_0$ $\gg$ [$\zeta_{CR}$ + $\zeta_{PHOTO}]/2k'$.
The models of \citet{draine1996} suggest that the \h2\ photodestruction rate will be 
comparable to the cosmic ray destruction rate adopted here for a column density
of \h2\ $\simeq$ $10^{21}$ cm$^{-2}$. 
Thus, while there could be a modest contribution from photodissociation to the
total \h2\ destruction rate, it will not be enough to change the time constant.
However, the photodestruction will increase the steady--state abundance of HI.  
The fact that the values we have measured are only modestly above those expected solely from
cosmic ray destruction of \h2\ indicates that the effect is modest.
If photodestruction plays a significant role, our
solutions become lower limits to $\tau_{cloud}$ since they would, in effect,
be steady--state values.

Third, as discussed in Section \ref{atmoldensities}, we may have underestimated the column
densities of molecular hydrogen, and hence n$_0$, along the lines of sight to the 
cores of these clouds.  
This may result in our having modestly  overestimated n$_{HI}$/n$_0$ in these directions.
This would have the effect of making the clouds look ``younger'' than they actually
are.  

Fourth, we must consider the possibility that there is not a reservoir of chemisorbed hydrogen atoms
on the grain as postulated by \citep{cazaux2002, cazaux2004}.  
Then, if the gas phase HI density and
thus the adsorption rate is sufficiently low, one can reach a situation in
which  $<$N$_{HI}$$>$, the average number of hydrogen atoms on the grain, is less than unity.
A newly--adsorbed incident atom may not find another atom with which to combine,
and the \h2\ formation rate will be reduced. 
This issue has been discussed by \citet{biham1998} and \citet{katz1999}, and is consistent
with laboratory measurements \citep[e.g.][]{vidali1998}.  
There is some concern whether the laboratory measurements are really applicable to
the surfaces of interstellar grains, so that it seems appropriate to be cautious
about the relevance of the measurements.  
If this picture is correct, the rate of \h2\ formation for small grains is
also reduced since $<$N$_{HI}$$>$ will also be less than unity.
These effects combine to reduce the overall \h2\ formation rate, particularly
at relatively late times in the cloud evolution, when n$_{HI}$ $\leq$ 100 cm$^{-3}$.

Additional support for our deteriming the minimum age of dark molecular clouds comes
from modeling of the thermal evolution of gas which is in the process of converting
hydrogen from atomic to molecular form.
Each hydrogen molecule formed releases a few eV of energy into the gas.  
The exact value is uncertain due to unknown distribution of the 4.5 eV molecular binding
energy between grain heating, internal energy of the H$_2$ (which is radiated
from the cloud), and kinetic energy of the H$_2$, which eventually is thermalized
\citep{flower1990}.  
Under steady state conditions and high densities as we have seen in Section 
\ref{timedep_sec}, the density of atomic hydrogen is $\simeq$ few cm$^{-3}$.
This density of HI results from the balance between formation of HI via 
cosmic ray destruction of H$_2$ and destruction of HI by formation of HI on grains.
In this situation, the heating from H$_2$ formation can be considered as a part
of the cosmic ray heating, as it is the latter which supplies the energy to 
break the molecular bonds.

Early in the evolution of a cloud from atomic to molecular form, if
we assume that the beginning of the process is characterized by a sharp increase
in density to something approximating its current value, then the heating rate
from H$_2$ formation is much larger, by a factor n$_0$/n$_{HI}^*$ $\simeq$ 10$^3$
\citep{goldsmith1978}.
The conversion of atomic to molecular hydrogen thus results in a reheating of
the cloud (which had been cooling after the initial compression), to a temperature
$\simeq$ 100 K \citep{flower1990}.  
This reheating phase lasts for a time approximately equal to the characteristic
HI $\rightarrow$ \h2\ conversion time (equation \ref{timeconst}), which is
plausibly on the order of a million years.
The fact that we see dark clouds which have obviously cooled to temperatures
an order of magnitude below that sustained by the initial formation of molecular
hydrogen is a further indication that the initiation of the molecular cloud phase
must have taken place a minimum of a few million years in the past.

\subsection{Other Effects and Concerns}

Although the time--dependent evolution of the HI fractional abundance in
the scenario developed above seems plausible, there are a number of other
processes which may affect the abundance of atomic hydrogen and thus confuse
this picture.  
One such effect is turbulent diffusion, in which the enhanced ability
of material to move over significant distances increases the effect of
concentration gradients \citep{xie1995}.  
Turbulent diffusion has been applied by \citet{willacy2002} to the case
of atomic hydrogen in dark clouds.  
Their results suggest that a quite modest values of the turbulent diffusion
constant, K $\leq$ $5\times10^{21}$ cm$^3$ s$^{-2}$ are sufficient to
produce atomic hydrogen densities in the centers of clouds similar to
what we measure.  
This is not surprising, in the sense that our values are equal to or modestly
greater than those expected for steady--state conditions with no turbulent
diffusion.  

It appears that while these values of K may be physically plausible, they
are significantly smaller than those required for this model to reproduced
various other chemical abundances in dark clouds.
It is also the case that the atomic hydrogen column densities predicted by
this model are significantly in excess of those that we observe.
Thus, it is premature to say that the present observations indicate, in
any significant way, that turbulent diffusion plays a very significant
role.  
The results of \citet{willacy2002} for the density of HI in the center of
the model cloud are for $5\times10^6$ yr of evolution of the cloud, and
it is plausible that the timescale we have derived above still is relevant
for reducing the HI density in the centers of dark clouds to the low values
we have obtained.

A related model suggests large--scale circulation of gas in the
cloud  \citep{chieze1989}.  In this model, parcels of
gas are exchanged between the outer, relatively unshielded portions of
the cloud, and the interior.  
These authors analyzed the effect on a number of molecular species, but did
not consider HI.  
However, it is reasonable that the effect would be to increase the HI density
inside the cloud, as is the case for other species which are typically abundant
only in regions of small visual extinction.
While suggestive, it is difficult to make any quantitative comparisons
with this model as no results for atomic hydrogen are reported.

We should consider the correlation between cold HI and molecular column density 
presented earlier in Figure \ref{hinsa13co}.  
It is evident that in the context of a steady--state for HI production and
destruction, the HI column  density is proportional to the line of sight extent
of the cloud, since the number density of atomic hydrogen is a constant.
In this sense it is very different than the column density of e.g. \th, which
is the integral of the number density of that species along the line of sight.
For the HI, the extent of the cloud is be defined by two factors.  
The first of these is the size of the region in which the HI number density 
is more or less constant. 
Evidently, the fractional abundance of the HI will increase dramatically in
the portion of the cloud with low extinction, but outside of some radius, 
the number density will become insignificant.
The second factor is basically observational: since the absorption coefficient
for HI varies inversely as the temperature, atomic hydrogen in the outer layers 
of the cloud is much less detectable than that in the cold interior.  
This effect is significant as the outer layers of the cloud even with standard
radiation field can be at temperatures $\simeq$ 100 K, an order of magnitude
higher than those in the cloud center.

It is appropriate here to follow up on this issue of warm HI surrounding that
which we feel is well--mixed with the molecular material.
Models of basically static clouds embedded in a much warmer, diffuse interstellar
medium and irradiated by an interstellar radiation field (ISRF) have envelopes (``onion
skins'') of ionized hydrogen, and neutral hydrogen, and ionized carbon, and neutral
carbon \citep[e.g.][]{abgrall1992, lebourlot1993}.  
The self--shielding of the H$_2$
results in its abundance rising when the visible optical depth $\tau_v$ reaches
10$^{-4}$, and becoming comparable to that of HI for $\tau_v$ = 10$^{-1}$.
The temperature will be 30 K and 100 K, for enhancement factors of the radiation
field relative to the standard ISRF of 1 and 10, respectively.  

The HI halos seen in emission around molecular clouds 
\citep[e.g.][]{wannier1983, andersson1992} correspond to these warm cloud edges
where the fractional abundance of H$_2$ is dropping significantly.
However, as discussed earlier, the very narrow line widths and low minimum line
intensities that we are dealing with here indicate that we are observing much
cooler and also less turbulent gas.
The smaller spatial extent of HINSA absorption compared to the CI emission as
well as its narrower line width (Section \ref{CI}) is additional strong evidence that 
the HINSA column density is not significantly affected by the warm neutral/ionized
cloud halos
	\footnote{This is the same conclusion reached by \cite{flynn2002} based 
	on modeling the HI self absorption produced by cold clouds in the Galactic disk.}
.

The cold HI and molecular column densities, as indicated by Figure \ref{hinsa13co} 
(as well as by the maps of their distributions), are quite well correlated.
Given the very different distributions expected from theoretical models,
this appears somewhat surprising, but may reflect to some degree the basic geometry
of the clouds. 
While there is clearly a significant density gradient, these regions have a basically convex
geometry in three dimensions, so that lines of sight with large column density
do encompass larger dimensions with sufficient material to provide a significant
contribution to the column of cold HI.
More complete models including the spatial and temporal evolution of atomic and molecular
hydrogen, as well as other species, will be necessary to assess how accurately
the atomic to molecular abundance ratio in the centers of these clouds can be used
to determine their age.

\subsection{Upper Limit on the Cosmic Ray Ionization Rate}

We can utilize our measurement of the atomic hydrogen density to set a useful upper
limit on the cosmic ray ionization rate in reasonably well--shielded dense clouds.  
From equation \ref{resid} we can write the cosmic ray ionization rate in terms of
the steady state atomic hydrogen density and the \h2\ formation rate as
\be
\zeta_{H_2} = 2k'_{H_2}n_{HI}^* \lp
\ee
The HI density we have measured is an upper limit to the steady state density of
atomic hydrogen, since processes such as turbulent diffusion, imperfect shielding
against photodissociation, and  time--dependent cloud evolution all work to increase
the atomic hydrogen density.  
As discussed in Section \ref{reh2form}, the uncertainties in k$'_{H_2}$
are substantial, but an upper limit to k$'_{H_2}$ equal
to 6$\times10^{-17}$ cm$^3$ s$^{-1}$ is reasonable.  
If we interpret our observations as giving n$_{HI}$ = 4 cm$^{-3}$ as
the upper limit to n$_{HI}^*$, we find that 
$\zeta_{H_2} \leq 5\times10^{-16}$ s$^{-1}$.

Our data, in common with many other observations of dark clouds, constrain the
cosmic ray flux through the heating provided to dark clouds.  
In regions with visual extinction greater than a few magitudes, cosmic rays
are expected to be the dominant heating source for the interstellar gas.  
At densities $\leq$ 10$^4$ cm$^{-3}$ gas--dust coupling has only a very
minor effect on the gas temperature, which is determined by the balance
between cosmic ray heating and molecular line cooling \citep{goldsmith2001}.
The latter varies as the gas temperature to the 2.4 power for n$_{H_2}$ = 10$^3$ 
cm$^{-3}$ and 2.7 power for n$_{H_2}$ = 10$^4$ cm$^{-3}$ 

The minimum gas temperature is obtained with no depletion of coolant species,
and for n$_{H_2}$ in the range 10$^3$ to 10$^4$ cm$^{-3}$, we find that
$T_{gas}$ = $T_0 g^{0.4}$, where $g$ is the scaling factor for the cosmic ray
heating rate relative to a reference value 10$^{-27}$ n$_{H_2}$ erg cm$^{-3}$ s$^{-1}$.
$T_0$ is equal to 10 K for n$_{H_2}$ = 10$^3$ cm$^{-3}$ and 13 K for 
n$_{H_2}$ = 10$^4$ cm$^{-3}$.

For our adopted cosmic ray ionization rate $\zeta$(H$_2$) = 5.2$\times$10$^{-17}$ s$^{-1}$,
and taking $\Delta$Q = 7 eV as the average energy transferred to the gas per
cosmic ray ionization in a region of low fractional ionization \citep{cravens1978}, 
we obtain a heating rate $\Gamma$ = 5.8$\times$10$^{-28}$n$_{H_2}$ erg cm$^{-3}$ s$^{-1}$.
This heating rate corresponds to g = 0.58, which yields gas temperatures 
$T_{gas}$ = 8.0 K for n$_{H_2}$ = 10$^3$ cm$^{-3}$ and 10.5 K for n$_{H_2}$ = 10$^4$ cm$^{-3}$.
These are in good agreement with the temperatures of clouds, 8 K to 10 K determined 
in many studies using \tw\ \citep[e.g.][] {li2003}, and ammonia \citep{tafalla2002}.
These temperatures are also in excellent agreement with the low central temperatures 
inferred from modeling preprotostellar cores \citep{evans2001}.

To be consistent with measured values of the kinetic temperature, we 
cannot have $g$ substantially greater than unity.
Taking a reasonable upper limit to be g = 1 yields $\Gamma_{cr}$ $\leq$ 
1$\times10^{-27}$n$_{H_2}$ erg cm$^{-3}$ s$^{-1}$, and gives gas temperatures 
$T_{gas}$ = 10 K for n$_{H_2}$ = 10$^3$ cm$^{-3}$ and 13 K for n$_{H_2}$ = 10$^4$ cm$^{-3}$.
A cosmic ray heating rate significantly greater than this value thus seems unlikely
to characterize the dense regions in dark clouds.
The much higher values that are sometimes inferred \citep[e.g.][]{mccall2003, liszt2003} may be 
the result of local conditions or of more diffuse compared to denser regions, 
since it is relatively more difficult to depress the heating rate than to enhance it.
The low temperature of molecular clouds such as those observed here also suggests
that the upper limit for the cosmic ray ionization
rate derived above may well be overly generous and that the
actual value of $\zeta_{H_2}$ does not significantly exceed 10$^{-16}$ s$^{-1}$.

\section{SUMMARY}

We have carried out a study of four dark interstellar clouds to define better the location of the
cold atomic hydrogen relative to the molecular component within these regions.
The HI is traced through observations of narrow self absorption features (HINSA), which
are seen against the general Galactic background.
The molecular clouds are traced by observations of OH (Arecibo) and \th\ and \ce\ obtained
at FCRAO.
For comparisons between atomic and molecular species, our results are limited to
regions $\geq$ 3$\arcmin$ in size, and thus refer to moderate--scale structure
of these clouds.  The most quiescent well--shielded central regions, in which
molecular depletion may be significant, are heavily beam--diluted in this study
and thus are unlikely to be affecting our conclusions.

The extent of the cold HI is close to, but somewhat smaller
than that of the \th, and larger than the size of the \ce~ emitting region.
This is consistent with results previously obtained by \citet{li2003}, in which
the nonthermal line widths of the species follow this same ordering.
The column density of cold HI is moderately well correlated with that of
\th, which is consistent with the general impressions of the maps of the
different species, and which further indicates that the atomic hydrogen we are
observing is not in outer envelopes of these clouds, but is predominantly
in well-shielded cold regions, again consistent with the low temperatures
for the HINSA--producing gas derived by \citet{li2003}.

The mapping of \th\ and \ce\ isotopologues reveals that the ratio of the column
density of these species rises sharply at the edges of the clouds.  This
is unlikely to be a result of any radiative transfer effect, but rather 
suggests that the abundance ratio is enhanced in regions of low extinction.
This effect is consistent with chemical isotopic fractionation, which has
been predicted to be significant where the abundance of ionized carbon is
significant, but the temperature is still quite low.  
The enhancement of the $^{13}$C in carbon monoxide offsets the drop in
the abundance of this molecule in the outer regions of clouds due to
photodestruction.
The result is that the size of the region over which the abundance of \th\ relative 
to \h2\ remains relatively uniform is increased, and we feel that this
isotopologue is to be preferred as a tracer of the large--scale cloud structure,
although saturation effects in regions of largest column density must be considered.

Using the HINSA and \th\ data, we model these clouds and obtain the central densities of
atomic hydrogen and \h2.  
The central densities of \h2 are 800 to $\simeq$ 3000 cm$^{-3}$,
and those of cold HI are between 2 and 6 cm$^{-3}$.
The ratio of the atomic hydrogen to total proton densities ranges from 0.5 to 
3 $\times10^{-3}$.  
The values of n$_{H_2}$ from \ce\ are a factor of two larger than those from \th,
and the fractional abundances of HI a corresponding factor lower.
These absolute and fractional HI abundances are close to, but somewhat larger than those
expected for steady--state balance between HI production by cosmic ray ionization
of \h2, and production of this molecule by grain surface recombination of HI.
The effect of a power law grain size distribution on the formation rate of \h2\ molecules has
been evaluated, and for a MRN distribution with reasonable upper and lower limits, we find
that that \h2\ production rate is increased by a factor of 3.4 over that for
standard grains of radius 1700 \AA.

A number of processes including mass exchange and turbulent diffusion may
be playing a role in determining the abundance of atomic hydrogen in dark clouds,
but it is not certain whether their contribution is significant.   
We assume that these clouds have been evolving at constant density since
the time that the atomic to molecular transition was initiated by some process
which suddently increased the visual extinction and eliminating photodestruction
of \h2.  
In this idealized picture, the HI to \h2 ratio drops monotonically with time,
and the time constant for the conversion is given by
$\tau_{HI\rightarrow H_2} = 2.7\times10^9/n_0$ yr, where n$_0$ is the total proton 
density.  
For the central densities of these sources, the time constants are $\simeq$ $10^6$ yr,
and the ages of these clouds are 3 to 10 million years.  
The relatively low fractional abundance of atomic hydrogen makes it extremely
unlikely that the time since the compressive event starting the atomic to molecular
transition is significantly less than this, since photodestruction and the
other processes mentioned above increase the HI abundance, and thereby require
that the grain surface production of \h2\ be quite close to completion.  
The conversion of atomic to molecular hydrogen sets a quite large
lower limit to the time scale for the ``molecularization'' of dense clouds, 
and thus for the overall process of star formation.

\acknowledgments

This work was supported in part by NASA through contract NAS5-30702 for
operation of SWAS, and by the National Astronomy and Ionosphere Center.
We thank Gianfranco Vidali and Xander Tielens for providing valuable
insights to their work on \h2\ formation on dust grains.  We appreciate
the information on cosmic ray heating and ionization provided by
Adam Burrows, Steve Doty, and Floris van der Tak.  Information on turbulence
and \h2\ formation was generously provided by Steve Stahler and David Hollenbach.
We thank the anonymous referees for a variety of helpful general and specific
comments.

\appendix
\section{Effect of Grain Size Distribution on \h2\ Formation Rate}

In order to investigate the effect of a grain size distribution, we start
from equation \ref{rh2single}.  
For a grain size
distribution, again assuming other quantities are independent of grain size,
we use equation \ref{KH2}, but must take the appropriate grain cross section
and grain volume averaged over the grain size distribution
\footnote{
The quantity which needs to explicitly be integrated over $a$ is the grain cross 
section.  Integrating the grain volume over $a$ is necessary only to normalize
by a fixed total mass in grains, given that we have assumed that all grains 
have the same density.}
.  
We can write the formation rate coefficient for a grain size distribution as
\be
\label{KH2dist}
k_{H_2} = \frac{S_{HI}\epsilon_{H_2}<v_{HI}><m_{gas}>}{2\rho_{gr}GDR} 
		\frac{\Sigma_{gr}}{V_{gr}}\lp
\ee

Assuming a distribution of spherical grains, and taking the relative number of 
grains having radius between $a$ and $a + da$ to be $n(a)da$, we 
have

\be
\label{avgsig}
\Sigma_{gr} = \int n(a)\sigma(a) da~,
\ee
and
\be
\label{avgvol}
V_{gr} = \int n(a) V(a) da~.
\ee

We assume a power law grain size distribution having the form
\be
n(a) = n(a_{min})(\frac{a}{a_{min}})^{-\alpha}~,
\ee
where $a_{min}$ is the minimum grain radius and $a_{max}$ is the
maximum grain radius ($a_{min} \leq a \leq a_{max}$). 
The power law exponent, $\alpha$, is 3.5 in MRN model, but is here left 
as a free parameter.
The total density of grains (cm$^{-3}$) is given by\footnote{
The dimensions of $n(a)$ and $n(a_{min})$ are evidently cm$^{-4}$.
}
\be
\label{ntot}
n_{tot} = \int n(a) da = \int^{a_{max}}_{a_{min}} n(a_{min})(\frac{a}{a_{min}})^{-\alpha} da~.
\ee
Evaluating equation \ref{ntot}, we find
\be
n(a_{min}) = \frac{1 - \alpha}{R^{1 - \alpha} - 1} \frac{n_{tot}}{a_{min}}~,
\ee
where $R$ is the ratio of maximum to minimum grain radius:
\be
R = \frac{a_{max}}{a_{min}}~.
\ee

The ratio of the integrated grain cross section to integrated grain volume is
\be
\label{sigmavgrain1}
\frac{\Sigma_{gr}}{V_{gr}} = 
			\frac{3}{4a_{min}}
			\left[\frac{4 - \alpha}{3 - \alpha}\right]
			\frac{R^{3 - \alpha} - 1}{R^{4 - \alpha} - 1}~.
\ee
The effect of grain size distributions characterized by different values of
$\alpha$ on $\Sigma_{gr}/V_{gr}$
is shown as a function of $R$ in Figure \ref{gsd_effect}.

In the limit of no grain size distribution (R $\rightarrow$ 1) we recover
the expected result for spherical grains of a single radius, $a_s$, 
\be
\left.\frac{\Sigma_{gr}}{V_{gr}} \right|_{a = a_s} = \frac{3}{4a_s}~.
\ee
We can consider various values of $a_{max}$ and $a_{min}$ by treating $a_{min}$
and $R$ as free parameters, but it is helpful to normalize to the
cross section to volume ratio for a single grain size, which gives us
\be
\label{sigmavgrain2}
\frac{\Sigma_{gr}}{V_{gr}}\left/\frac{3}{4a_s}\right. = 
			\frac{a_s}{a_{min}}\left[\frac{4 - \alpha}{3 - \alpha}\right]
			\frac{R^{3 - \alpha} - 1}{R^{4 - \alpha} - 1}~.
\ee
Equations \ref{sigmavgrain1} and \ref{sigmavgrain2}
are general, but of particular relevance to suggested astronomical grain size
distributions, these expressions are well--behaved for all values of $\alpha 
\ge 0$, including e.g. 3 and 4.

\clearpage
\begin{figure}[ht]
\plotone{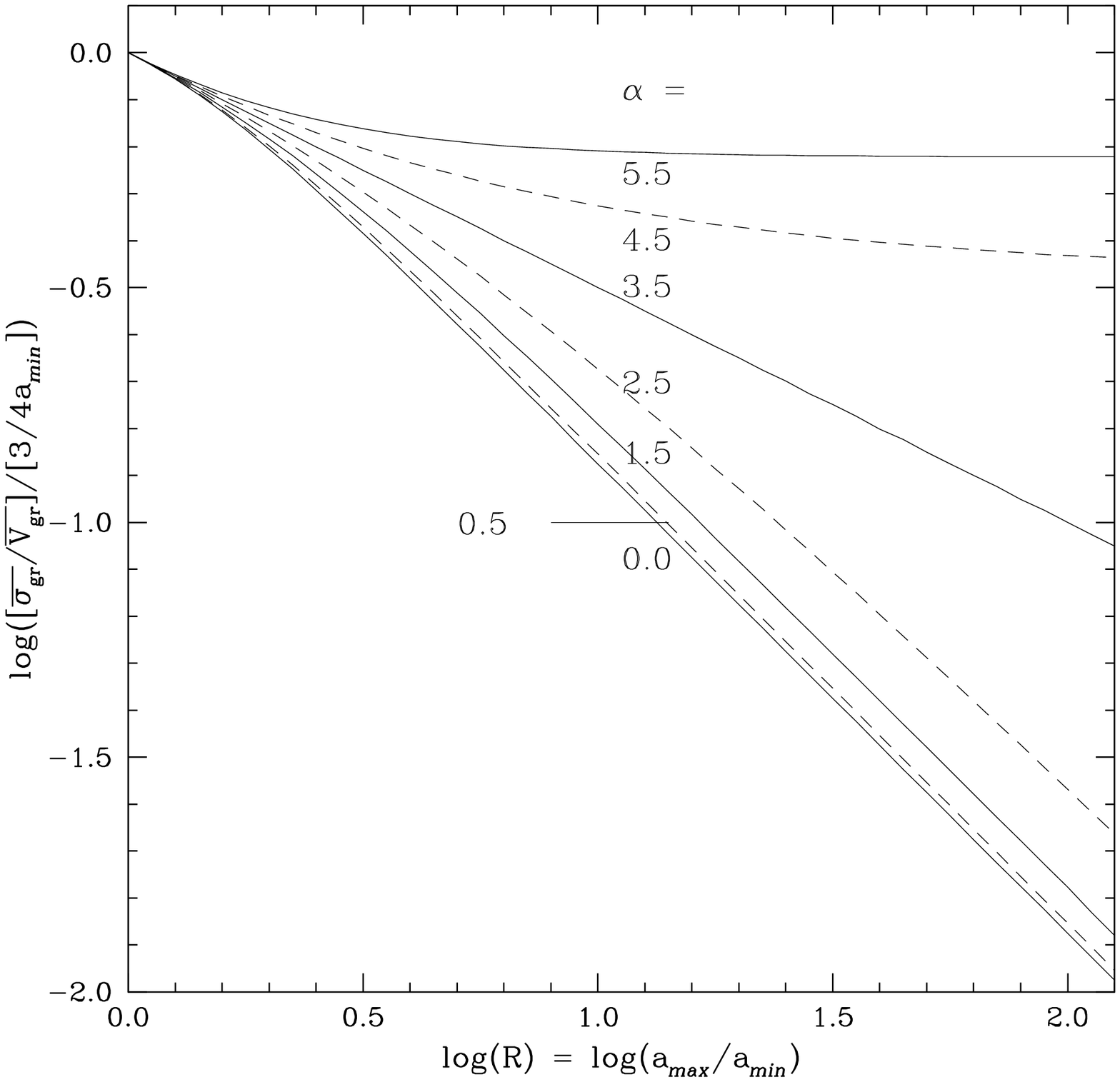}
\caption{\label{gsd_effect}Effect of grain size distribution with 
$n(a) \propto a^{-\alpha}$ on
the ratio of the integrated grain cross section to the integrated grain volume.  The H$_2$
formation rate is directly proportional to this ratio.  
The vertical axis is normalized to the value of 
$\Sigma_{gr}/V_{gr}$ for grains of a single radius $a_{min}$.}
\end{figure}
\clearpage

For a fixed value of $a_{min}$, $\Sigma_{gr}/V_{gr}$ 
decreases as $R$ increases.  For $R$ $\gg$ 1, the ratio varies as $R^{-1}$ 
for 0 $\leq$ $\alpha$ $\leq$ 2.5, and has a logarithmic dependence for 
$R$ = 3.0 and 4.0. 
The MRN grain size distribution is characterized by $\alpha$ = 3.5. 
In this case, we find that, for all $R$,

\be
\label{mrn1}
\left.\frac
{\Sigma_{gr}}{V_{gr}}\left/\frac{3}{4a_{min}}\right.
\right|_{\alpha = 3.5} =
			R^{-0.5}~.
\ee
This particularly simple result facilitates assessing the impact of
this grain size distribution. 
The unnormalized ratio of the integrated grain cross section to integrated grain  
volume can be written
\be
	\label{mrn-unnorm}
	\left.\frac
	{\Sigma_{gr}}{V_{gr}}
	\right|_{\alpha = 3.5} =
	\frac{3}{4} \frac{1}{\sqrt{a_{max} a_{min}}}~,
\ee
indicating that for this value of $\alpha$, the \h2\ formation rate is 
inversely proportional to the geometric mean of the extremes of the 
grain size distribution.

In order to interpret the effect of the grain size distribution quantitatively,
we need to consider specific values of the ``standard'' single grain size,
$a_s$, along with the specifics of the GSD defined by $a_{min}$ and $R$,
in addition to $\alpha$.  
Normalizing relative to the single--sized grains, we find
\be
	\label{mrn2}
	\left.
	\frac{\Sigma_{gr}}{V_{gr}}\left/\frac{3}{4a_{s}}\right.
	\right|_{\alpha = 3.5} =
			\frac{a_s}{a_{min}}R^{-0.5} = 
			\frac{a_s}{\sqrt{a_{max}a_{min}}}~.
\ee
The ratio of integrated grain cross section to integrated grain volume relative
to the ratio for grains of a single radius is thus determined by the ratio of
that single grain radius to the geometric mean of the maximum and minimum
grain radii.  
A smaller grain has a larger surface to volume ratio,
so a grain size distribution including grains that are all smaller than
the standard grain, will have $\Sigma_{gr}/V_{gr}$
larger than $3/4a_{s}$.
It is more meaningful to compare grain size distributions with 
the constraint $a_{min}$ $\leq$ $a_s$ $\leq$ $a_{max}$, so that
for $R \rightarrow 1$, $(\Sigma_{gr}/V_{gr})/(3/4a_s)$ 
$\rightarrow a_s/a_{min} = 1$.
Note that as long as $a_{min}$ $\leq$ $a_s$ $\leq$ $a_{max}$, the
normalized grain cross section per unit grain volume is independent
of the reference grain size $a_s$.

There is a vast and ever--expanding literature on the subject of the
range of grain sizes in interstellar clouds.  
These should be considered relative to the ``standard'' dust grain
size, which is based on fitting behavior of the extinction in the
visible region of the spectrum \citep[][ch 7]{spitzer1978}. 
A representative value of $a_s$ is 1700 \AA.

The more diverse dust grain population includes grains having radii 
as small as 10 \AA~ which are transiently heated to $\simeq$ 10$^3$ K  
in reflection nebulae by absorption of a single photon \citep{sell84}.
The infrared emission of small grains is also seen from the edges of
molecular clouds exposed to fairly standard interstellar radiation field
\citep{beichman88}.
The need for small grains (radius 5--20 \AA) to explain diffuse infrared
emission in a number of galaxies in addition to the Milky Way was
emphasized by \citet{rowan92}.
Specific spectral features seen in the infrared led to recognition that
the population of small grains may include polycyclic aromatic hydrocarbon
(PAH) molecules \citep*[e.g.~][]{leger84, omont86, ryter87, puget89}, 
although very small grains may well have a significant graphite component 
\citep{desert86}.

There has been increasing evidence that a larger--than--standard grain size 
component exists in dense regions of the interstellar medium. 
The earliest indications of ``grain growth'' in these regions came
from visual/infrared photometry \citep{carrasco73} and the wavelength
dependence of polarization \citep*[][]{carrasco73,brown75,
serkowski75,whittet78,breger81}.
Large grains (radius $\simeq$ 5000 \AA) have been invoked to reproduce the
spectral properties of reflection nebulae \citep{pendleton90} and
30 \mic~(3$\times$10$^5$ \AA) radius grains added to a dust model to 
explain the emission observed at millimeter wavelengths \citep{rowan92}.

Grain growth has also been inferred from the dependence of the long--wavelength 
(millimeter and submillimeter) spectral index of the dust emissivity
on the column density and volume density in molecular clouds  \citep*[][]{woody89,
walker90,goldsmith97,visser98}.  These observations indicate that the 
frequency dependence (characterized by power
law exponent $\beta$) of the grain emissivity or optical 
depth in dense regions of molecular clouds has $\beta$ less 
than 2, the standard value characteristic of grains in the general 
interstellar medium \citep*[e.g.][]{kogut96}.
The observations of \citet{stepnik2003} suggest that grain coagulation in a dense
filament in the Taurus complex has resulted in a deficiency of small grains
together with an enhancement of the submillimeter wavelength emissivity.
Grain coagulation is a very complex and imperfectly understood process (see
Draine 1985 and references therein), but can result in fluffy
and highly non--spherical grains \citep{wright1987}.  The MRN
grain size distribution may be significantly modified in regions where 
dsut coagulation has taken place, with a resulting uncertain effect on 
the \h2\ formation rate.

Evidence for grain growth is very strong in even denser regions, particularly
protostellar disks, which suggests that this is, in fact, a continuous
process throughout all phases of the dense interstellar medium.  
In choosing nominal values for the maximum and minum grain sizes for
use in evaluating the \h2\ formation rate in fairly standard clouds,
there is certainly a large range of possible values, but the effect
of the uncertainty is mitigated by the fact that in e.g. equation \ref{mrn2}
the dependence is on the geometric mean of $a_{max}$ and $a_{min}$. 
We adopt nominal values for radii $a_{max}$ = 10,000 \AA\ and $a_{min}$ = 25 \AA,
which together with the the standard grain radius $a_s$ = 1,700 \AA\ 
result in the grain size distribution increasing the \h2\ formation rate
by a factor of 3.4.  This modest but significant value suggests that a
plausible grain size distribution does increase the \h2\ formation rate, but
only in extreme cases where the grain size distribution has been heavily
skewed to either very large or very small grains, will the formation
rate coefficient $k'_{H_2}$ be drastically affected by the grain size
distribution.


\section{Potential Energy of Cloud with Gaussian Density Distribution}
\label{gaussvirial}

If we take the density distribution for a spherically symmetric
cloud to be given by
\be
n(r) = n_0 e^{-(r/r_0)^2}\lc
\ee
the mass contained within radius $r$ is given by
\be
M(r) = \pi r_0^3 [\sqrt{\pi}~erfc(\frac{r}{r_0}) -2(\frac{r}{r_0})e^{-(\frac{r}{r_0})^2}]\lc
\ee
where $erfc$ is the error function.
The total mass of the cloud is then given by
\be
M = \pi^{3/2}n_0<m_{gas}>r_0^3 \lc
\ee
where $<m_{gas}>$ is the average mass of a gas particle, defined in equation \ref{mgasavg}.

The potential energy is given by the integral
\be
\U = - G\int_0^\infty \frac{M(r)dm(r)}{r}\lc
\ee
where $dm(r)$ is the mass in a shell at radius $r$.
The integral is well--defined, and gives us the result
\be
\U = - \frac{1}{\sqrt{2\pi}} \frac{GM^2}{r_0}\lp
\ee
A fraction 0.90 of the mass is contained within radius $r_{90}$ = $1.768r_0$, and expressed
in terms of this parameter, $\U = -0.71GM^2/r_{90}$, very similar to that for a $r^{-1}$
density distribution.



\clearpage
\begin{deluxetable}{lllcl}
\tablewidth{0pt}
\tablecaption{\label{sources} Source Parameters}
\tablehead{
	 \colhead{Name} 	& \colhead{RA(2000)} 	& \colhead{Decl (2000)} 
	&\colhead{Distance} 	& \colhead{Reference}\\
	 \colhead{}		& \colhead{}		& \colhead{}
	&\colhead{(pc)}		& \colhead{}
	}
\startdata
L1544	& 05 04 18.1	& 25 11 08	& 140	&  \citet{elias1978}\\
B227	& 06 07 28.4	& 19 28 04	& 400	&  \citet{bok1974}; \citet{arquilla1984}\\
	&		&		&	&  \citet{tomita1979}\\
L1574	& 06 08 05.0	& 18 28	12	& 300	&  \citet{kawamura1998} \\
CB45	& 06 08 45.9	& 17 53 15	& 300	&  \citet{kawamura1998}\\
\enddata
\end{deluxetable}


\begin{deluxetable}{llll}
\tablewidth{200pt}
\tablecaption{\label{tpeak} Peak T$_{mb}$ (K) at Reference Positions of Sources}
\tablehead{
	\colhead{Name} 	& \colhead{OH}	& \colhead{\th}	& \colhead{\ce}}
\startdata
L1544(0,0)  \tablenotemark{a}		&  1.5	& 5.9	& 2.4 	\\
B227(0,0)				&  0.36	& 2.6	& 0.41	\\
L1574(0,9)  \tablenotemark{b}		&  0.46	& 2.3	& 0.44	\\
CB45(-3,18) \tablenotemark{c}		&  0.44	& 3.4	& 0.79	\\
\enddata
\tablenotetext{a}{offsets in arcminutes}
\tablenotetext{b}{peak of blue emission}
\tablenotetext{c}{stronger peak in the cloud}
\end{deluxetable}


\begin{deluxetable}{lllllll}
\tablewidth{0pt}
\tablecaption{\label{hioh1318} Column Densities at Reference Positions}
\tablehead{
	  \colhead{Name} 		& \colhead{N(HINSA)}		&\colhead{N(OH)}
	 &\colhead{N(\th)}		& \colhead{N(\ce)}		&\colhead{N(\h2)\tablenotemark{a}}
	 &\colhead{N(\h2)\tablenotemark{b}}\\
	  \colhead{}			& \colhead{10$^{19}$ cm$^{-2}$}	
	 &\colhead{10$^{14}$ cm$^{-2}$} & \colhead{10$^{14}$ cm$^{-2}$}	
	 &\colhead{10$^{14}$ cm$^{-2}$}	& \colhead{10$^{21}$ cm$^{-2}$}	
	 & \colhead{10$^{21}$ cm$^{-2}$}
	  }
\startdata
L1544		&\phn.42	&1.7		&70\tablenotemark{c}.	&12.	&6.0	&8.4\\
B227		&\phn.61	&\phn.80	&31.			&\phn2.6&3.2	&2.8\\
L1574r (0,0)	&\phn.83	&2.3		&\phn9.4		&\nodata&1.7	&\nodata\\
L1574b (0,9)	&1.5		&2.1		&36.			&\phn5.1&3.6	&4.3\\
CB45 (-3,18)	&2.7		&1.0		&62.			&\phn6.8&5.4	&5.3\\
\enddata
\tablenotetext{a}{From \th\ (column 4) and expression on p 603 of \citet{frerking1982} for \th\ in Taurus}
\tablenotetext{b}{From \ce\ (column 5) and expression on p 603 of \citet{frerking1982} for \ce\ in Taurus}
\tablenotetext{c}{Corrected for saturation as described in the text}
\end{deluxetable}
	%
	%

\begin{deluxetable}{lllll}
\tablewidth{0pt}
\tablecaption{\label{densities} Cloud Dimensions and Central Densities of Different Species}
\tablehead{	 \colhead{Name} 				&\colhead{Species}	
		&\colhead{$\Delta\theta_{FWHM}$\tablenotemark{a}}	
		&\colhead{$\Delta$z$_{FWHM~LOS}$\tablenotemark{a}}		&\colhead{n$_{cen}$}\\
		 \colhead{}					&\colhead{}		&\colhead{(\arcmin)}	
		&\colhead{(10$^{18}$ cm)} 	&\colhead{(cm$^{-3}$)}
		 }

\startdata
L1544~	&HINSA	&$>$11.5		&$>$1.3			&$<$3.0\\
L1544~	&\th	&$>$14.3		&$>$1.8			&$<$3.7$\times10^{-3}$\\
L1544~	&\ce	&\phm{$>$}\phn9.3	&\phm{$>$}1.2		&\phm{$>$}9.4$\times10^{-4}$\\
\\
B227~	&HINSA	&\phm{$>$}\phn8.1	&\phm{$>$}2.9		&\phm{$>$}2.0\\	
B227~	&\th	&\phm{$>$}\phn8.4	&\phm{$>$}3.0		&\phm{$>$}9.6$\times10^{-4}$\\
B227~	&\ce	&\phm{$>$}\phn5.2	&\phm{$>$}1.9		&\phm{$>$}1.3$\times10^{-4}$\\
\\
L1574r~	&HINSA	&\phm{$>$}\phn7.4	&\phm{$>$}2.0		&\phm{$>$}3.9\\
L1574r~	&\th	&\phm{$>$}\phn7.7	&\phm{$>$}2.1		&\phm{$>$}4.2$\times10^{-4}$\\
\\
L1574b~	&HINSA	&\phm{$>$}\phn8.8	&\phm{$>$}2.4		&\phm{$>$}5.9\\
L1574b~	&\th	&\phm{$>$}11.2		&\phm{$>$}3.0		&\phm{$>$}1.1$\times10^{-3}$\\
L1574b~	&\ce	&\phm{$>$}\phn6.2	&\phm{$>$}1.7		&\phm{$>$}2.8$\times10^{-4}$\\
\\
CB45~	&HINSA	&\phm{$>$}15.6		&\phm{$>$}4.2		&\phm{$>$}6.0\\
CB45~	&\th	&\phm{$>$}12.0 		&\phm{$>$}3.0		&\phm{$>$}1.9$\times10^{-3}$\\
CB45~	&\ce	&\phm{$>$}10.0		&\phm{$>$}2.7		&\phm{$>$}2.4$\times10^{-4}$\\
\enddata
\tablenotetext{a}{Geometric mean of two dimensions in the plane of the sky}
\end{deluxetable}


\begin{deluxetable}{llllll}
\tablewidth{0pt}
\tablecaption{\label{atmoldens} Central Atomic\tablenotemark{a}~~and Molecular Hydrogen Densities}
\tablehead{	 \colhead{Name} 		
		&\colhead{n$_{HI}$}	
	  	&\colhead{n$_{H_2}$ from \th\tablenotemark{b}} 
	  	&\colhead{n$_{H_2}$ from \ce\tablenotemark{c}}	
	  	&\colhead{n$_{HI}$/n$_0$    \tablenotemark{d}}
	  	&\colhead{n$_{HI}$/n$_0$}\\
					
	   	 \colhead{ }			
	   	&\colhead{(cm$^{-3}$)}	
	  	&\colhead{(cm$^{-3}$)} 	
	  	&\colhead{(cm$^{-3}$)}	
	  	&\colhead{from \th}			
	  	&\colhead{from \ce}
          }
\startdata

L1544	&$<$3.0		&$<$3100		&\phm{$<$}6400		&$4.8\times10^{-4}$\tablenotemark{e}	&$2.3\times10^{-4}$\\
B227	&\phm{$<$}2.0	&\phm{$<$}1000 		&\phm{$<$}1200		&$1.1\times10^{-3}$	&$8.3\times10^{-4}$\\	
L1574r	&\phm{$<$}3.9	&\phm{$<$}\phn800	&\phm{$<$}\nodata	&$2.4\times10^{-3}$	&\nodata	   \\
L1574b	&\phm{$<$}5.9	&\phm{$<$}1100		&\phm{$<$}2400		&$2.7\times10^{-3}$	&$1.2\times10^{-3}$\\
CB45	&\phm{$<$}6.0	&\phm{$<$}1700		&\phm{$<$}1900		&$1.8\times10^{-3}$	&$1.6\times10^{-3}$\\

\enddata
\tablenotetext{a}{Cold HI seen in absorption (HINSA)}
\tablenotetext{b}{From \h2\ column density derived from N(\th) given in Table \ref{hioh1318}, column 6}
\tablenotetext{c}{From \h2\ column density derived from N(\ce) given in Table \ref{hioh1318}, column 7}
\tablenotetext{d}{Total proton density in molecular and atomic hydrogen}
\tablenotetext{e}{Taking the source dimensions equal to upper limits measured}
\end{deluxetable}


\begin{deluxetable}{llllll}
\tablewidth{0pt}
\tablecaption{\label{virialparam} Cloud Masses and Energies}
\tablehead{	 \colhead{Name} 		
		&\colhead{$\Delta v_{FWHM}$ \tablenotemark{a}}	
	  	&\colhead{$\sigma_0$	\tablenotemark{b}} 
	  	&\colhead{M		\tablenotemark{c}}	
	  	&\colhead{$2\T$}
	  	&\colhead{$-\U$ \tablenotemark{d}}\\
					
	   	 \colhead{ }			
	   	&\colhead{(km s$^{-1}$)}	
	  	&\colhead{($10^4$ cm s$^{-1}$)} 	
	  	&\colhead{($10^{35}$ g)}	
	  	&\colhead{($10^{45}$ erg)}			
	  	&\colhead{($10^{45}$ erg)}\\
          }
\startdata

L1544	&0.64	&\phn4.6&\phn4.8	&\phn1.5	&\phn5.7\tablenotemark{e}\\
B227	&1.08	&\phn7.9&\phn6.7	&\phn4.8	&\phn6.6\\	
L1574r	&1.42	&10.4	&\phn1.9	&\phn2.3	&\phn0.76\\
L1574b	&1.43	&10.5	&\phn7.6	&\phn9.1	&\phn8.5\\
CB45	&1.50	&11.0	&13.		&17.		&25.\\

\enddata
\tablenotetext{a}{Full width to half maximum line width from \th}
\tablenotetext{b}{3 dimensional velocity dispersion defined by 
		$\sigma_0 = [3(\frac{\Delta v_{FWHM}^2}{8ln2} - \frac{kT}{m})]^{0.5}$,
		where the line width and mass are those for \th}
\tablenotetext{c}{Based on Gaussian density distribution and central densities and 
		dimesions given in Table \ref{densities} and including a correction for helium}
\tablenotetext{d}{Calculated using Gaussian density distribution; see Appendix \ref{gaussvirial}}
\tablenotetext{e}{Assuming that we have measured total mass, and that actual dimension is defined by
		our lower limit}
\end{deluxetable}

\end{document}